\documentclass[journal]{IEEEtran}

\usepackage{url}
\usepackage{./bs_macros}
\usepackage[labelformat=simple]{subcaption}

\graphicspath{{./Figures/}}

\theoremstyle{definition}

\newtheorem{remark}{Remark}

\newtheorem{example}{Example}

\makeatletter
\patchcmd{\@maketitle}
{\addvspace{0.5\baselineskip}\egroup}
{\addvspace{-1\baselineskip}\egroup}
{}
{}
\makeatother

\begin{document}



\title{A Probabilistic Spectral Analysis of Multivariate Real-Valued Nonstationary Signals}

\author{
	
Bruno Scalzo, \IEEEmembership{Student Member, IEEE}, Ljubi$\check{\text{s}}$a Stankovi\'c, \IEEEmembership{Fellow, IEEE}, Danilo P. Mandic, \IEEEmembership{Fellow, IEEE}
	
\thanks{B. Scalzo and D. P. Mandic  are with the Department of Electrical and Electronic Engineering, Imperial College London, London SW7 2AZ, U.K. (e-mail: bruno.scalzo-dees12@imperial.ac.uk; d.mandic@imperial.ac.uk).}
\thanks{L. Stankovi\'c is with the Faculty of Electrical Engineering, University of Montenegro, D$\check{\text{z}}$ord$\check{\text{z}}$a Va$\check{\text{s}}$ingtona bb, 81000 Podgorica, Montenegro, e-mail: ljubisa@ucg.ac.me.}


}


\maketitle

\begin{abstract}
A class of multivariate spectral representations for real-valued nonstationary random variables is introduced, which is characterised by a general complex Gaussian distribution. In this way, the temporal signal properties -- harmonicity, wide-sense stationarity and cyclostationarity -- are designated respectively by the mean, Hermitian variance and pseudo-variance of the associated time-frequency representation (TFR). For rigour, the estimators of the TFR distribution parameters are derived within a maximum likelihood framework and are shown to be statistically consistent, owing to the statistical identifiability of the proposed distribution parametrization. By virtue of the assumed probabilistic model, a generalised likelihood ratio test (GLRT) for nonstationarity detection is also proposed. Intuitive examples demonstrate the utility of the derived probabilistic framework for spectral analysis in low-SNR environments.
\end{abstract}


\IEEEpeerreviewmaketitle

\section{Introduction}

Spectral estimation is at the heart of exploratory data analysis and is an inherently complex-valued task. Yet, the results are commonly interpreted using magnitude-only based models, thus not accounting for the information within the phase spectrum. From the maximum entropy viewpoint, magnitude-only models make the implicit, yet fundamental, assumption that the phase information, which is intrinsic to complex-valued spectral data, is uniform and thus not informative. More formally, such an assumption implies that the probability density function (pdf) of a spectral variable is complex \textit{circular}, or \textit{rotation-invariant}, in the complex plane \cite{Picinbono1994}.

While it was first noted in \cite{Picinbono1994,Edelblute1996} that wide-sense stationary (WSS) random signals exhibit circularly distributed Fourier transforms, the existence of noncircularly distributed counterparts was not considered until the seminal work in \cite{Picinbono1996}. Thereafter, a class of general complex time-frequency representations (TFRs) was proposed in \cite{Schreier2003}, which has motivated various developments of complex-valued spectral analysis techniques for modelling real-valued nonstationary random signals \cite{Schreier2008_2, Millioz2010, Clark2012_2, Mandic2014, Mandic2017, Scalzo_c_panorama_2018, Wisdom2015, Wisdom2016, Okopal2015, Walden2018}. These methods employ \textit{augmented complex statistics} \cite{Naseer1993}, which provide a complete second-order statistical description of a general complex random variable, $x \in \domC$, by incorporating the standard Hermitian variance, $\expect{|x|^{2}} \in \domR$, which parametrizes the scale of a pdf, in conjunction with the pseudo-variance, $\expect{x^{2}} \in \domC$, which parametrizes the eccentricity and angle of an elliptical pdf in the complex plane \cite{Ollila2008}. If the pseudo-variance is equal to zero, $\expect{x^{2}}=0$, then a complex variable, $x$, is said to be \textit{proper}, or \textit{second-order circular}. Otherwise, it is \textit{improper} or \textit{second-order noncircular}, which implies a rotation-dependent pdf and manifests itself in unequal powers or non-zero degree of correlation between the real and imaginary parts of $x$ \cite{Mandic2009}.


In the spectral analysis setting, existing techniques typically consider the spectral expansion of a real-valued signal, $x(t) \in \domR$, of the form
\begin{equation}
	x(t) = \int_{-\infty}^{\infty} e^{\jmath \omega t} X(\omega) \, d\omega \label{eq:spectral_expansion_intro}
\end{equation}
where the complex-valued spectral variable, $X(\omega) \in \domC$, exhibits two distinctive second-order moments,
\begin{equation}
	R(\omega) = \expect{|X(\omega)|^{2}}, \quad P(\omega) = \expect{X^{2}(\omega)} \label{eq:PSD_CSD_intro}
\end{equation}
These are referred to respectively as the \textit{power spectral density} and \textit{complementary spectral density} \cite{Schreier2003} (or \textit{panorama} \cite{Mandic2014}). The condition of \textit{spectral noncircularity}, whereby $|P(\omega)|>0$, has been demonstrated to be related to deterministic phase characteristics of a signal in the time domain \cite{Clark2012_2}. This highlights that the consideration of the \textit{augmented spectral statistics} becomes indispensable in the analysis of nonstationary signals.

Despite success, there remain several issues that need to be addressed for a more widespread application of noncircular spectral variables and their augmented statistics:
\begin{enumerate}[label=\arabic*),leftmargin=4mm]
	\item Observe that the augmented spectral statistics in (\ref{eq:PSD_CSD_intro}) are \textit{absolute}, or \textit{non-centred}, moments. Referring to the maximum entropy principle, if no assumptions are made on the mean of a spectral process, $X(\omega)$, then the implicit, yet fundamental, assumption of pure stochasticity is imposed on its time-domain counterpart, $x(t)$. As such, existing techniques cannot distinguish between the deterministic and stochastic properties exhibited by real-world signals.
	\item Spectral estimation algorithms operate under the assumptions of \textit{ergodicity}, whereby the statistical expectation (ensemble-average) and time-average operators are equivalent. However, nonstationary signals have non-uniformly distributed phase in time and are therefore \textit{non-ergodic}.
	\item Existing spectral analysis frameworks for real-valued signals are typically derived within the univariate setting only, and there is a need to extend these to cater for the multivariate setting, to support many emerging applications.
\end{enumerate}
To this end, we introduce a class of multivariate complex time-frequency representations which exhibits ergodicity in the first- and second-order moments. These conditions facilitate the statistical analysis of nonstationary time-domain signals even in the critical case where only a single realisation is observed. The proposed parametrization of TFR distributions, based on the mean and second-order central moments, is shown to allow for a rigorous and unifying treatment of a general class of multivariate nonstationary real-valued temporal signals which describe both deterministic and stochastic behaviours, a typical case in real-world signals.


The rest of this paper is organized as follows. Section \ref{sec:max_entropy} provides a maximum entropy viewpoint of existing spectral estimation techniques and motivates the need for general complex Gaussian distributed TFRs, which are shown to describe a broad class of nonstationary time-domain signals in Section \ref{sec:class_nonstationary_processes}. A compact formulation of the spectral statistics is derived in Section \ref{sec:augmented_spectrum} to enable the derivation of their estimators in Section \ref{sec:MLE}. Lastly, a GLRT detector for nonstationarity is proposed in Section \ref{sec:GLRT}. Numerical examples are provided and demonstrate the advantages of the proposed probabilistic framework for spectral analysis even in low-SNR environments.



\section{Multivariate Complex TFR}

\subsection{Notation}

We employ the following notation, whereby the Fourier transform of a function, $x(t) \in \domR$, is given by a complex spectral representation, $X(\omega) \in \domC$, that is
\begin{equation}
x(t) \FourierArrow X(\omega) \label{eq:Fourier_transform_scalar}
\end{equation}
Similarly, the Fourier transform of a vector-valued function, $\x(t) \in \domR^{N}$, is given by $\bbx(\omega) \in \domC^{N}$, that is
\begin{equation}
\x(t) \FourierArrow \bbx(\omega)
\end{equation}
The above mapping implies the element-wise mapping in (\ref{eq:Fourier_transform_scalar}).

For convenience, and with a slight abuse of notation, the absolute value, denoted by the symbol $|\cdot|$, and trigonometric operators, denoted by $\angle(\cdot)$ and $\cos(\cdot)$, are employed in an element-wise manner when the argument is vector- or matrix-valued. Otherwise, the symbol $\odot$ is employed to denote the Hadamard (element-wise) product.

\subsection{Maximum entropy spectral representations}

\label{sec:max_entropy}

We begin by demonstrating that standard spectral models implicitly impose the assumption of pure temporal stochastic signals and hence fail to model deterministic temporal properties. To see this, consider the spectral expansion of a signal, $x(t) \in \domR$, in (\ref{eq:spectral_expansion_intro}). Typically, our \textit{a priori} knowledge of a signal is limited to the following statistical conditions:
\begin{enumerate}[label=\roman*)]
	\item \textit{Square integrability}, 
	\begin{equation}
		\int_{-\infty}^{\infty} x^{2}(t) \, dt < \infty
	\end{equation}
	\item  \textit{Mean-square ergodicity} (Parseval's theorem), 
	\begin{equation}
		\int_{-\infty}^{\infty} x^{2}(t)  \, dt = \int_{-\infty}^{\infty} |X(\omega)|^{2} \, d \omega
	\end{equation}
\end{enumerate}
	In such situations of limited knowledge, it is natural to resort to the \textit{maximum entropy principle} \cite{Jaynes1957}, which states that the most representative random process maximises entropy given the currently available knowledge. From this viewpoint, the stochastic model which best represents the spectral process, $X(\omega) \in \domC$, is a zero-mean circularly distributed Gaussian random variable \cite{Burg1967,Cover1984}, which exhibits the following first- and second-order moments
\begin{equation}
	 \expect{X(\omega)} \! = \! 0, \quad \!\!\! \expect{X^{2}(\omega)} \! = \! 0, \quad \!\!\! \expect{|X(\omega)|^{2}} \! = \! R(\omega)
\end{equation}
Upon inspecting the moments of the time-domain counterpart,
\begin{align}
	\expect{x(t)} & =  \int_{-\infty}^{\infty} e^{\jmath \omega t} \expect{X(\omega)} \, d\omega = 0 \\
	\var{x(t)} & = \int_{-\infty}^{\infty} \expect{|X(\omega)|^{2}} \, d\omega =  \int_{-\infty}^{\infty} R(\omega) \, d\omega
\end{align}
we find that the time-domain signal represents a zero-mean Gaussian distributed random variable, given by
\begin{equation}
	x \sim \Normal{0,\int_{-\infty}^{\infty} R(\omega) \, d\omega}
\end{equation}
In other words, in classical spectral analysis the assumption of \textit{pure temporal stochasticity} is implicitly imposed. To address this issue, it is necessary to assert statistical assumptions on the \textit{mean} of the spectral representation, for instance, that $M(\omega)=\expect{X(\omega)} \neq 0$. In this way, we obtain the following deterministic (harmonic) signal
\begin{equation}
	m(t) = \expect{x(t)} = \int_{-\infty}^{\infty} e^{\jmath \omega t} M(\omega) \, d\omega \label{eq:harmonic_intro}
\end{equation}
The benefits of employing such statistical assumptions will be investigated in this work.




\subsection{Multivariate complex TFR with ergodicity in the moments}

\label{section:moments}

In practice, we typically deal with single realisations of finite duration; we can therefore assume that the signal under consideration is of finite power and has a well-defined time-average, which implies respectively the conditions of \textit{square} ($L_{2}$-norm) and \textit{absolute} ($L_{1}$-norm) \textit{integrability}, given by
\begin{align}
	\lim_{T \to \infty} \frac{1}{T} \int_{-\frac{T}{2}}^{\frac{T}{2}} \|\x(t)\|_{2} \, dt & < \infty \\
	\lim_{T \to \infty} \frac{1}{T} \int_{-\frac{T}{2}}^{\frac{T}{2}} \|\x(t)\|_{1} \, dt & < \infty
\end{align}
Under these conditions, the real-valued signal, $\x(t) \in \domR^{N}$, admits the following time-frequency expansion \cite{Loeve1977}
\begin{equation}
	\label{eq:TFR_expansion}
	\x(t) = \int_{-\infty}^{\infty} e^{\jmath \omega t} \bbx(t,\omega) \, d \omega
\end{equation}
where $\bbx(t,\omega) \in \domC^{N} $ is the realisation of a spectral process at an angular frequency, $\omega$, and time instant, $t$.

To ensure that $\x(t)$ is real-valued, the Hermitian symmetry condition, $\bbx^{\ast}(t,\omega) = \bbx(t,-\omega)$, must be satisfied so that we can express (\ref{eq:TFR_expansion}) in terms of TFRs associated with the positive angular frequencies only, to yield
\begin{equation}
	\x(t) = \int_{0}^{\infty} \left[ e^{\jmath \omega t}\bbx(t,\omega) + e^{- \jmath \omega t}\bbx^{\ast}(t,\omega) \right] \, d\omega \label{eq:TFR_expansion_positive}
\end{equation}

\subsection{Augmented spectral statistics}

To cater for a broad variety of deterministic and stochastic time-domain signals, the TFR is assumed to be multivariate general complex Gaussian distributed \cite{vandenBos1995}, i.e. $\bbx(t,\omega)$ follows the linear model
\begin{equation}
	\bbx(t,\omega) = \bbm(\omega) + \bbs(t,\omega)
\end{equation}
where $\bbs(t,\omega) \in \domC^{N}$ is a zero-mean stochastic process, while the time-invariant \textit{spectral mean} is given by,
\begin{equation}
	\expect{\bbx(t,\omega)} = \bbm(\omega) \in \domC^{N} \label{eq:spectral_mean}
\end{equation}
In addition, the time-invariant \textit{spectral covariance} and \textit{spectral pseudo-covariance} are respectively defined as
\begin{alignat}{2}
	\cov{\bbx(t,\omega)} & = \expect{\bbs(t,\omega)\bbs^{\Her}(t,\omega)} && = \bbR(\omega) \label{eq:spectral_covariance} \\
	\pcov{\bbx(t,\omega)} & = \expect{\bbs(t,\omega)\bbs^{\Trans}(t,\omega)} && = \bbP(\omega) \label{eq:spectral_pseudocovariance}
\end{alignat}
where the bound $\|\bbP(\omega)\|_{2} \leq \|\bbR(\omega)\|_{2}$ holds, by virtue of the Cauchy-Schwarz inequality.

As with multivariate complex variables in general, the TFR admits a compact \textit{augmented representation} of the form
\begin{equation}
	\ubbx(t,\omega) = \left[ \begin{array}{c}
		\bbx(t,\omega)	 \\
		\bbx^{\ast}(t,\omega)
	\end{array} \right] \in \domC^{2N}
\end{equation}
which can be used to compactly parametrize the pdf of $\bbx(t,\omega)$ as follows \cite{vandenBos1995}
\begin{equation}
	p(\ubbx,t,\omega) \! = \! \frac{ \exp \! \left[ \! - \frac{1}{2} \! \left( \ubbx(t,\omega) \! - \! \ubbm(\omega) \right)^{\Her} \! \ubbR^{-1} \!(\omega) \!  \left( \ubbx(t,\omega) \! - \! \ubbm(\omega) \right) \! \right] }{\pi^{N} \det^{\frac{1}{2}} ( \ubbR(\omega) )}
\end{equation}
with
\begin{alignat}{2}
	\ubbm(\omega) & = \expect{\ubbx(t,\omega)} && = \left[ \begin{array}{c}
		\bbm(\omega)	 \\
		\bbm^{\ast}(\omega)
	\end{array} \right] \\
	\ubbR(\omega) & = \cov{\ubbx(t,\omega)} && = \left[ \begin{array}{cc}
		\bbR(\omega) & \bbP(\omega) \\
		\bbP^{\ast}(\omega) & \bbR^{\ast}(\omega)
	\end{array} \right]
\end{alignat} 
being respectively the augmented spectral mean and covariance. Therefore, $\bbx(t,\omega)$ is said to be distributed according to
\begin{equation}
	\ubbx(t,\omega) \sim \CNormal{\ubbm(\omega),\ubbR(\omega)}
\end{equation}
Furthermore, if the time-frequency representations exhibit \textit{non-orthogonal bin-to-bin increments}, then it is necessary to also consider the following \textit{dual-frequency statistics} (for $\omega \neq \nu$)
\begin{align}
	\!\! \cov{\bbx(t,\omega),\bbx(t,\nu)} \! = \! \expect{\bbs(t,\omega)\bbs^{\Her}(t,\nu)} &= \bbR(\omega,\nu) \label{eq:dual_frequency_cov} \\
	\!\! \pcov{\bbx(t,\omega),\bbx(t,\nu)} \! = \! \expect{\bbs(t,\omega)\bbs^{\Trans}(t,\nu)} &= \bbP(\omega,\nu) \label{eq:dual_frequency_pcov}
\end{align}
which are respectively referred to as the \textit{dual-frequency spectral covariance} and \textit{dual-frequency spectral pseudo-covariance}, and exhibit the following properties
\begin{align}
	\bbR(\omega,\nu) & = \bbR^{\ast}(\nu,\omega)\\
	\bbP(\omega,\nu) & = \bbP(\nu,\omega)\\
	\|\bbP(\omega,\nu)\|_{2} \leq \|\bbR(\omega,\nu)\|_{2} & \leq \|\bbR(\omega)\|_{2}\|\bbR(\nu)\|_{2}
\end{align}
owing to the Cauchy-Schwarz inequality \cite{Schreier2003}.


\begin{remark} \label{remark:PSD}
	Notice that the spectral moments in (\ref{eq:spectral_mean})-(\ref{eq:spectral_pseudocovariance}) are \textit{centred}, which contrasts the usual spectral statistics based on the \textit{absolute} or \textit{non-centred} moments. It is therefore possible to express the standard power spectral density and and complementary spectral density \cite{Schreier2003,Schreier2008_2} (or panorama \cite{Mandic2014,Mandic2017,Scalzo_c_panorama_2018}), denoted respectively by $\tilde{\bbR}(\omega)$ and $\tilde{\bbP}(\omega)$, in terms of the spectral mean and covariances as follows
	\begin{align}
		\tilde{\bbR}(\omega) & = \expect{\bbx(t,\omega)\bbx^{\Her}(t,\omega)} = \bbm(\omega)\bbm^{\Her}(\omega) + \bbR(\omega) \label{eq:PSD} \\
		\tilde{\bbP}(\omega) & = \expect{\bbx(t,\omega)\bbx^{\Trans}(t,\omega)} = \bbm(\omega)\bbm^{\Trans}(\omega) + \bbP(\omega) \label{eq:CSD}
	\end{align}
	This shows that the mean and covariance information become entangled when employing the legacy absolute (non-centred) spectral statistics. This result also highlights that the power spectrum is inadequate for detecting harmonics in low signal-to-noise ratio environments, since $\|\bbR(\omega)\| \gg \|\bbm(\omega)\|^{2}$. Recall from (\ref{eq:harmonic_intro}) that harmonics are also parametrized by the spectral mean, $\bbm(\omega)$, while Gaussian noise is parametrized by the covariance, $\bbR(\omega)$. The power spectral density of the harmonics, $\tilde{\bbR}(\omega)$, would therefore be dominated by the power associated with the noise, thereby rendering the harmonic indistinguishable from the noise.
\end{remark}

\subsection{Temporal moments}

To gain insight into the physical meaning of the spectral moments, it is useful to re-express these statistics directly in the time domain by means of the inverse Fourier transform. For instance, the spectral mean exhibits the following relationship
\begin{align}
\bbm(\omega) & = \expect{\bbx(t,\omega)} = \int_{-\infty}^{\infty} e^{-\jmath \omega t}\expect{\x(t)}dt \notag\\
& = \int_{-\infty}^{\infty} e^{-\jmath \omega t}\m(t)dt
\end{align}
where
\begin{equation}
	\m(t) = \expect{\x(t)} \label{eq:TV_mean} 
\end{equation}
is the \textit{time-varying ensemble mean} of $\x(t)$. 

Upon introducing the zero-mean (centred) time-domain process, $\s(t) = \x(t) - \m(t)$, from (\ref{eq:dual_frequency_cov})--(\ref{eq:dual_frequency_pcov}) we can express the dual-frequency spectral covariances as follows
\begin{align}
\bbR(\omega,\nu) & = \expect{\bbs(t,\omega)\bbs^{\Her}(t,\nu)} \notag\\
& = \int_{-\infty}^{\infty}\int_{-\infty}^{\infty} e^{-\jmath (\omega t_{1} - \nu t_{2})} \expect{\s(t_{1})\s^{\Trans}(t_{2})} \, dt_{1}dt_{2} \notag\\
& = \int_{-\infty}^{\infty}\int_{-\infty}^{\infty} e^{-\jmath (\omega t_{1} - \nu t_{2})} \R(t_{1},t_{2}) \, dt_{1}dt_{2}
\end{align}
\begin{align}
\bbP(\omega,\nu) & = \expect{\bbs(t,\omega)\bbs^{\Trans}(t,\nu)} \notag\\
& = \int_{-\infty}^{\infty}\int_{-\infty}^{\infty} e^{-\jmath (\omega t_{1} + \nu t_{2})} \expect{\s(t_{1})\s^{\Trans}(-t_{2})} \, dt_{1}dt_{2} \notag\\
& = \int_{-\infty}^{\infty}\int_{-\infty}^{\infty} e^{-\jmath (\omega t_{1} + \nu t_{2})} \P(t_{1},t_{2}) \, dt_{1}dt_{2}
\end{align}
to obtain the second-order temporal statistics
\begin{alignat}{1}
\R(t_{1},t_{2}) & = \expect{\s(t_{1})\s^{\Trans}(t_{2})} \label{eq:centred_autocovariance} \\
\P(t_{1},t_{2}) & = \expect{\s(t_{1})\s^{\Trans}(-t_{2})} \label{eq:centred_autoconvolution}
\end{alignat}
which we refer to respectively as the ensemble \textit{autocovariance} and \textit{autoconvolution functions} of $\x(t)$. 

\begin{remark}
	The requirement of employing the complementary statistics to fully describe the complex spectral variables has a physically meaningful interpretation in the time domain. This is equivalent to exploiting the information arising from the \textit{time-symmetry} of $\x(t)$, which is captured by the autoconvolution in (\ref{eq:centred_autoconvolution}). This property was first investigated for univariate signals in \cite{Mandic2014}.
\end{remark}

\begin{remark} \label{remark:ergodicity}
	Observe that the time-domain statistics in (\ref{eq:TV_mean}) and (\ref{eq:centred_autocovariance})-(\ref{eq:centred_autoconvolution}) are defined using the expectation (ensemble-average) operator. Therefore, for general nonstationary signals with non-uniformly distributed phase in time, these statistics cannot be evaluated using the time-average operator owing to their \textit{non-ergodicity}. However, by virtue of the proposed framework, their TFRs, $\bbx(t,\omega)$, do exhibit ergodicity in the time-frequency domain. We demonstrate in the sequel that these statistics can be computed, even in the critical case where a single realisation of $\x(t)$ is observed.
\end{remark}


\section{A Class of Nonstationary Signals with Time-Invariant Spectral Statistics}

\label{sec:class_nonstationary_processes}

We now introduce a class of multivariate real-valued nonstationary temporal signals which exhibits the time-invariant spectral statistics introduced in Section \ref{section:moments}. 

Owing to the linearity property of the Fourier transform in (\ref{eq:TFR_expansion}), it follows that if the TFRs are multivariate complex Gaussian distributed, that is, $\ubbx(t,\omega) \sim \CNormal{\ubbm(\omega),\ubbR(\omega)}$, then their time-domain counterpart, $\x(t)$, is also multivariate Gaussian distributed, since a linear function of Gaussian random variables is also Gaussian distributed. The signal, $\x(t)$, is therefore distributed according to
\begin{equation}
\x(t) \sim \Normal{\m(t),\R(t)} \label{eq:pdf_time-varying_Gaussian}
\end{equation}
where $\m(t) \in \domR^{N}$ and $\R(t) \in \domR^{N \times N}$ are the time-varying mean vector and covariance matrix, defined respectively as
\begin{align}
	\m(t) & = \expect{\x(t)} \label{eq:time-varying_mean} \\
	\R(t) & = \cov{\x(t)} \label{eq:time-varying_cov}
\end{align}
which are a function of the spectral statistics, as is shown next.

\begin{remark}
	Observe that the time-varying statistics of $\x(t)$ in (\ref{eq:time-varying_mean})-(\ref{eq:time-varying_cov}) are defined using the expectation operator, and therefore cannot be estimated using the time-average operator in the time domain. Following from Remark \ref{remark:ergodicity}, we show in the sequel that it is possible to instead estimate time-varying statistics in the time-frequency domain, where the ergodicity condition applies.
\end{remark}

\subsection{Harmonic signal}

From (\ref{eq:time-varying_mean}), consider the statistical expectation of the spectral expansion of $\x(t)$ as in (\ref{eq:TFR_expansion}), to yield
\begin{equation}
\m(t) \! = \! \expect{\x(t)} \! = \!\! \int_{-\infty}^{\infty} \!\!\! e^{\jmath \omega t} \expect{\bbx(t,\omega)} d\omega \!  = \!\! \int_{-\infty}^{\infty} \!\!\! e^{\jmath \omega t} \bbm(\omega) d\omega \label{eq:spectral_expansion_mean}
\end{equation}
Alternatively, we can employ the equivalent formulation
\begin{equation}
	\m(t) = \int_{0}^{\infty} \cos\left( \omega t + \, \angle \bbm(\omega)  \right) \odot |\bbm(\omega)| \; d\omega \label{eq:spectral_expansion_mean_elementwise}
\end{equation}
Therefore, the time-varying mean of $\x(t)$ is also a multivariate real-valued \textit{harmonic} signal. Notice that for $\omega = 0$ the signal reduces to a multivariate DC component.

\begin{remark}
	Based on (\ref{eq:spectral_expansion_mean})--(\ref{eq:spectral_expansion_mean_elementwise}), observe that harmonics are solely parametrized by the spectral mean. Therefore, the task of estimating or detecting harmonics reduces to a \textit{first-order} statistical technique, whereby only the spectral mean needs to be considered. This contrasts the legacy spectral methods which employ absolute second-order statistics, such as the autocorrelation matrix, the power spectrum or the complementary spectrum, to estimate harmonics (see also Remark \ref{remark:PSD}).
\end{remark}

\begin{figure}[ht]
	\centering
	\includegraphics[width=0.24\textwidth, trim={0.2cm 0.2cm 0 0.2cm}, clip]{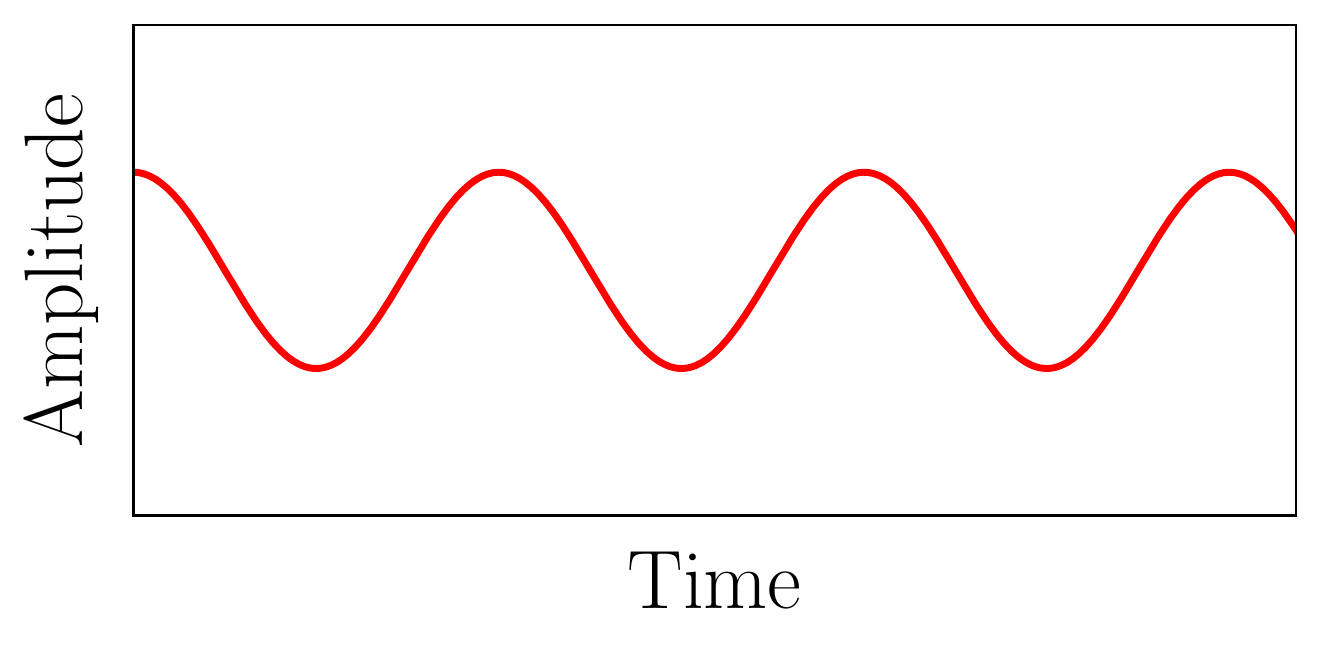}
	\hfill 
	\includegraphics[width=0.24\textwidth, trim={0.2cm 0.2cm 0 0.2cm}, clip]{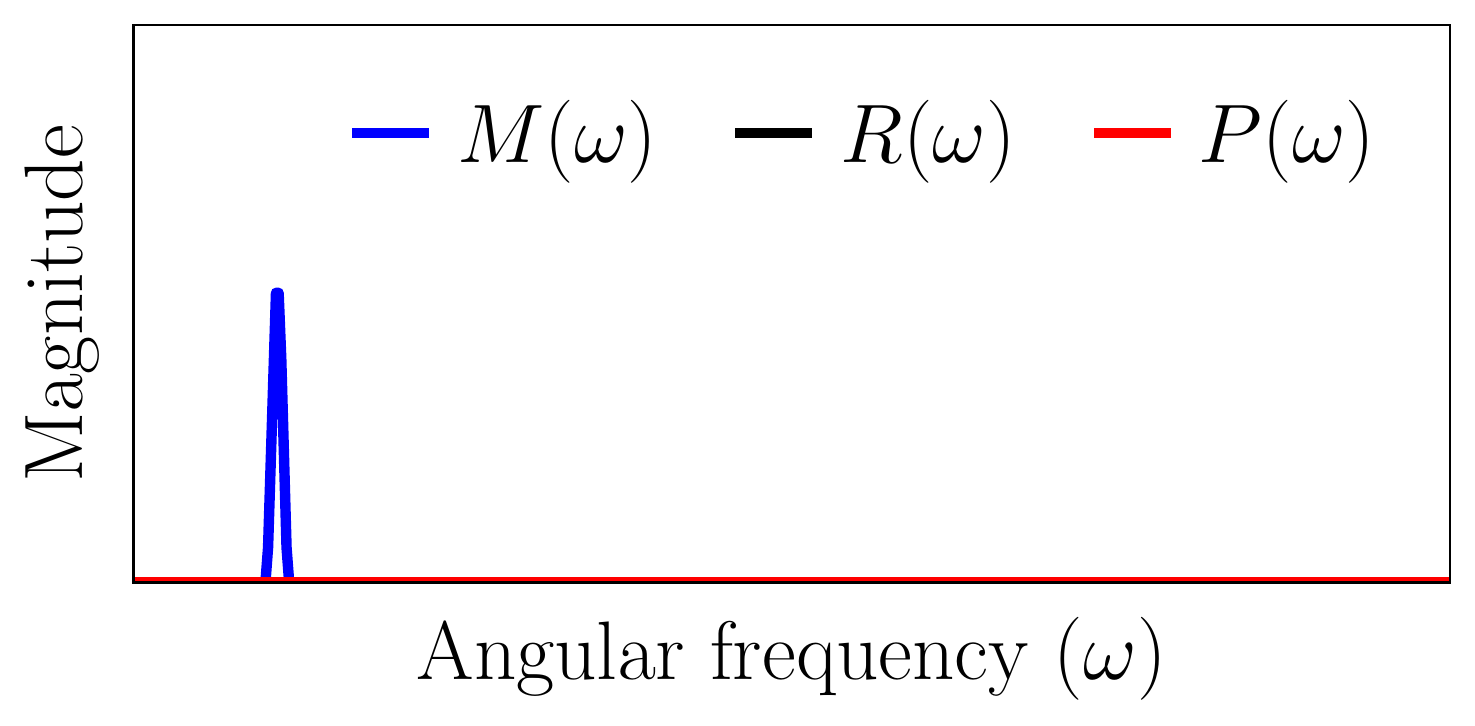} 
	\caption{\small A single realisation of a harmonic signal (left panel) and its associated spectral moments (right panel).}
	\label{fig:sine}
\end{figure}

\begin{example}
	For illustration purposes, Fig. \ref{fig:sine} shows a single realisation of a univariate \textit{harmonic} signal, denoted by $x(t) \sim \Normal{m(t),0}$, which exhibits a non-zero spectral mean at an angular frequency $\omega_{0} \neq 0$, that is, $|M(\omega_{0})|>0$. The time-varying mean is thus given by
	\begin{equation}
		m(t) = \cos\left( \omega_{0} t + \, \angle M(\omega_{0})  \right) |M(\omega_{0})|
	\end{equation}
\end{example}

\vspace{-0.4cm}

\subsection{General cyclostationary signal}

Following from the relation in (\ref{eq:time-varying_cov}), and upon introducing the centred signal, $\s(t) = \x(t)-\m(t)$, consider the covariance of the spectral expansion of $\x(t)$ as in (\ref{eq:TFR_expansion}), to obtain
\begin{align}
& \R(t) = \cov{\x(t)} = \expect{\s(t)\s^{\Trans}(t)} \notag\\
& = \int_{-\infty}^{\infty}\int_{-\infty}^{\infty} \left[ e^{\jmath (\omega-\nu) t} \bbR(\omega,\nu) +  e^{\jmath (\omega+\nu) t} \bbP(\omega,\nu) \right] \; d\omega d\nu \notag\\
& = \int_{0}^{\infty}\int_{0}^{\infty} \biggl[ \cos\left( (\omega-\nu) t + \angle \bbR(\omega,\nu) \right) \odot |\bbR(\omega,\nu)| \notag\\
& \quad + \cos\left( (\omega+\nu) t + \angle \bbP(\omega,\nu) \right) \odot |\bbP(\omega,\nu)| \biggr] \; d\omega d\nu \label{eq:general_cyclostationary_cov}
\end{align}
Therefore, the time-varying covariance of $\x(t)$ consists of a sum of \textit{general cyclostationary} components, each modulated at an angular frequency, $\omega$. 

\begin{remark}
	The class of real-valued cyclostationary signals introduced by the seminal work in \cite{Gardner1983,Gardner1986,Gardner1991_2} is a special case of the proposed class parametrized by the covariance in (\ref{eq:general_cyclostationary_cov}), whereby only the spectral Hermitian covariances, $\bbR(\omega,\nu)$, are considered. On the other hand, the model proposed in this work also considers the spectral complementary covariances, $\bbP(\omega,\nu)$, and can therefore cater for a more general class of nonstationary signals, as shown next.
\end{remark}

\begin{example}
	Fig. \ref{fig:semi_cyclo} shows a single realisation of a univariate \textit{general cyclostationary} signal, $x(t) \sim \Normal{0,r(t)}$, with a non-zero spectral Hermitian variance and pseudo-variance, at an angular frequency $\omega_{0} \neq 0$, such that, $|R(\omega_{0})|\geq|P(\omega_{0})|>0$. The signal therefore has a time-varying variance of the form
	\begin{equation}
		r(t) = R(\omega_{0}) + \cos\left( 2\omega_{0} t + \angle P(\omega_{0}) \right) |P(\omega_{0})| 
	\end{equation}
	
	\vspace{-0.4cm}
	
	\begin{figure}[ht]
		\centering
		\includegraphics[width=0.24\textwidth, trim={0.2cm 0.2cm 0 0.2cm}, clip]{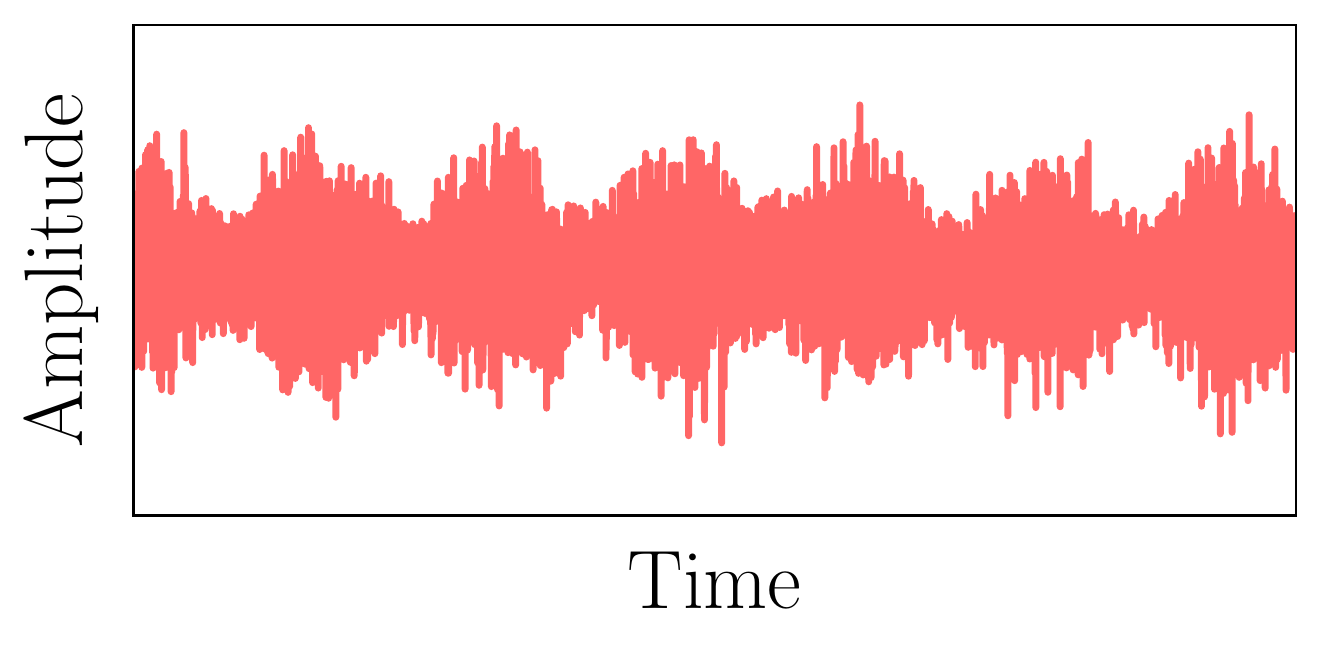}
		\hfill 
		\includegraphics[width=0.24\textwidth, trim={0.2cm 0.2cm 0 0.2cm}, clip]{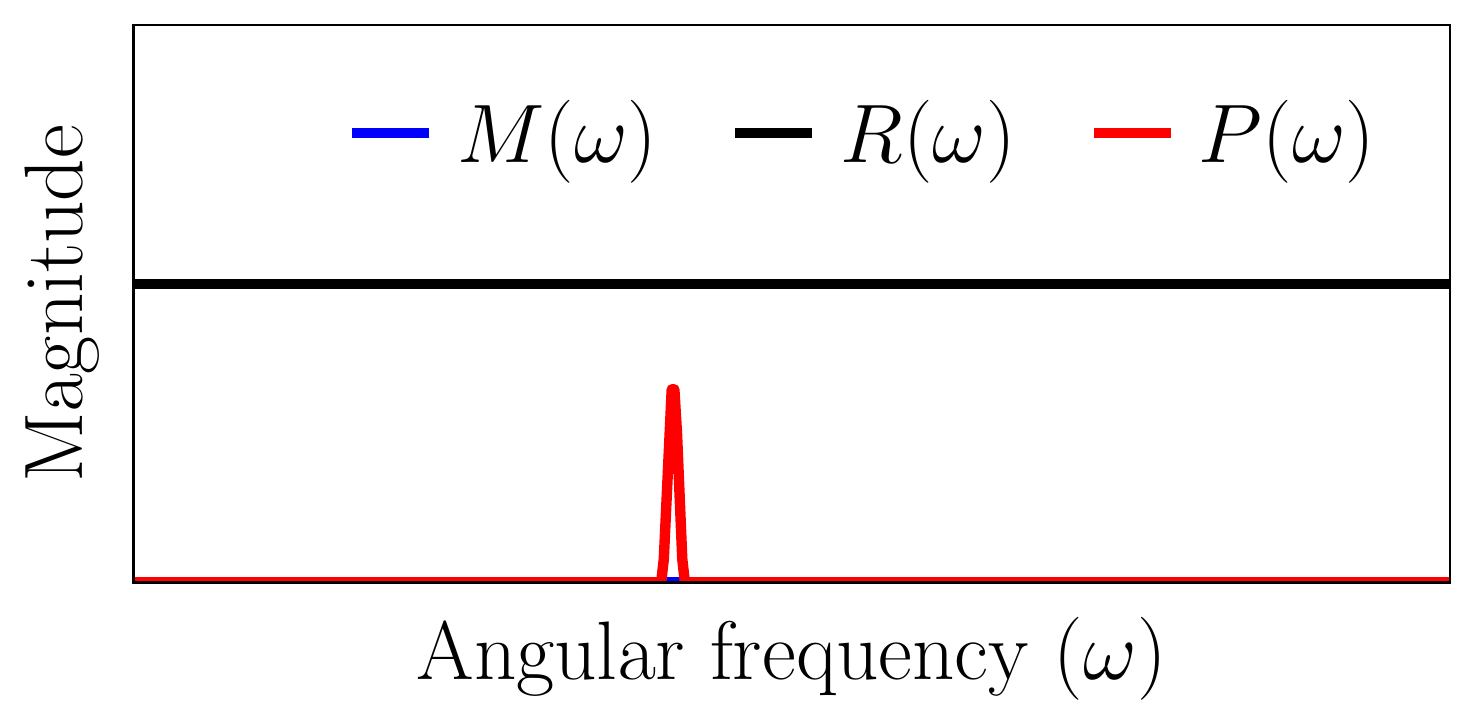} 
		\caption{\small A single realisation of a general cyclostationary signal (left panel) and its associated spectral moments (right panel).}
		\label{fig:semi_cyclo}
	\end{figure}
\end{example}

\noindent Next, consider the following special cases of (\ref{eq:general_cyclostationary_cov}):

\subsubsection{Wide-sense stationary signal}

Consider the case where the spectral Hermitian covariance is non-zero only on the stationary manifold ($\omega = \nu$), that is, $\bbR(\omega,\nu)=\0$ for $\omega \neq \nu$, and the TFR is circularly distributed, with $\bbP(\omega,\nu)=\0$ for all $\omega,\nu$. Then, the stationary signal, $\s(t)$, is wide-sense stationary. To see this, observe that the general cyclostationary covariance in (\ref{eq:general_cyclostationary_cov}) reduces to the time-invariant covariance 
\begin{equation}
\R(t) = \int_{0}^{\infty} \cos\left( \angle \bbR(\omega) \right) \odot |\bbR(\omega)| \; d\omega
\end{equation}

%
%
%
%


\begin{example}
	Fig. \ref{fig:WSS} displays a single realisation of a univariate \textit{wide-sense stationary} signal with non-zero spectral Hermitian variance at all frequencies, that is, $R(\omega)>0$ and $P(\omega)=0$ for all $\omega$. The variance of the signal is time-invariant, since
	\begin{equation}
		r(t) = \int_{-\infty}^{\infty }\cos\left( \angle R(\omega) \right) R(\omega) \, d\omega 
	\end{equation}

	\vspace{-0.4cm}

	\begin{figure}[ht]
		\centering
		\includegraphics[width=0.24\textwidth, trim={0.2cm 0.2cm 0 0.2cm}, clip]{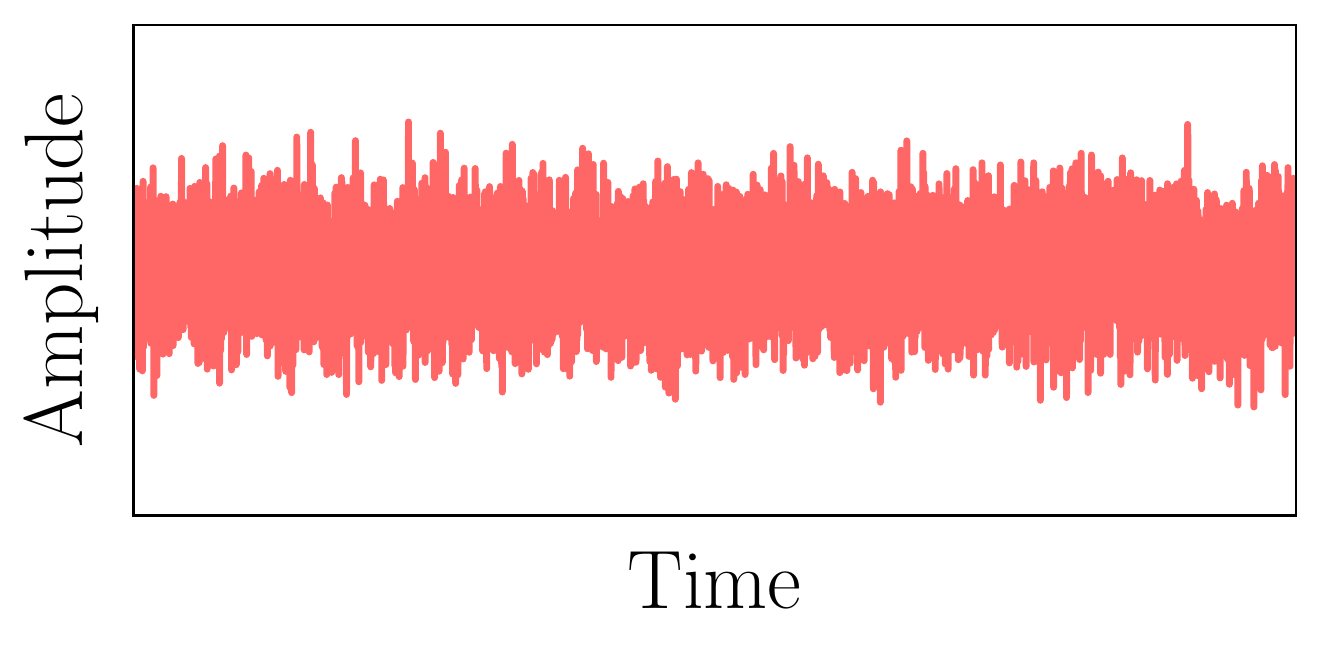}
		\hfill 
		\includegraphics[width=0.24\textwidth, trim={0.2cm 0.2cm 0 0.2cm}, clip]{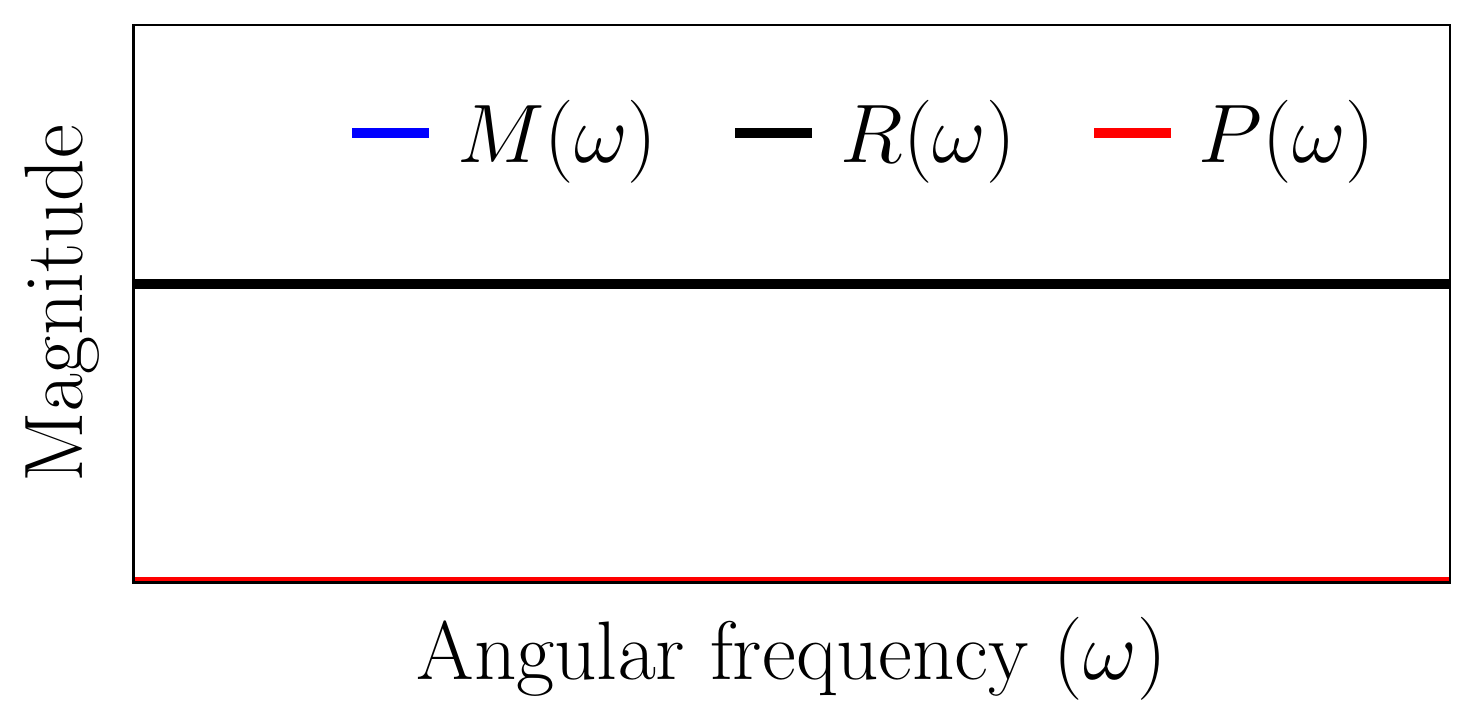} 
		\caption{\small A single realisation of a wide-sense stationary signal (left panel) and its associated spectral moments (right panel).}
		\label{fig:WSS}
	\end{figure}
	
\end{example}


\subsubsection{Pure cyclostationary process}

Consider the case where the TFRs are \textit{maximally noncircular}, or \textit{rectilinear}, whereby $\|\bbP(\omega,\nu)\|_{2}=\|\bbR(\omega,\nu)\|_{2}$. Using well-known trigonometric identities, the general cyclostationary covariance in (\ref{eq:general_cyclostationary_cov}) reduces to its purely cyclostationary counterpart, given by
\begin{equation}
	\R(t) \! = \!\! \int_{0}^{\infty}\!\!\!\!\int_{0}^{\infty}\!\!\!\!\! \cos\left( \omega t \! + \! \A_{\omega\nu} \right)\!\odot\!\cos\left( \nu t \! + \! \B_{\omega\nu} \right)\! \odot \! |\bbR(\omega,\nu)| d\omega d\nu
\end{equation}
with \vspace{-0.2cm}
\begin{align}
	\A_{\omega\nu} & = \frac{1}{2}\biggl( \angle \bbP(\omega,\nu)+\angle \bbR(\omega,\nu) \biggr) \\
	\B_{\omega\nu} & = \frac{1}{2}\biggl( \angle \bbP(\omega,\nu)-\angle \bbR(\omega,\nu) \biggr)
\end{align}

\begin{example}
	Fig. \ref{fig:pure_cyclo} illustrates a single realisation of a univariate \textit{purely cyclostationary} signal with non-zero spectral Hermitian variance and pseudo-variance at an angular frequency $\omega_{0}$, whereby $R(\omega_{0})=|P(\omega_{0})|>0$. The signal therefore has a time-varying variance of the form
	\begin{equation}
		r(t) = \cos^{2}\left( \omega_{0} t + \frac{1}{2} \angle P(\omega_{0}) \right) R(\omega_{0})
	\end{equation}

	\vspace{-0.4cm}

	\begin{figure}[ht]
		\centering
		\includegraphics[width=0.24\textwidth, trim={0.2cm 0.2cm 0 0.2cm}, clip]{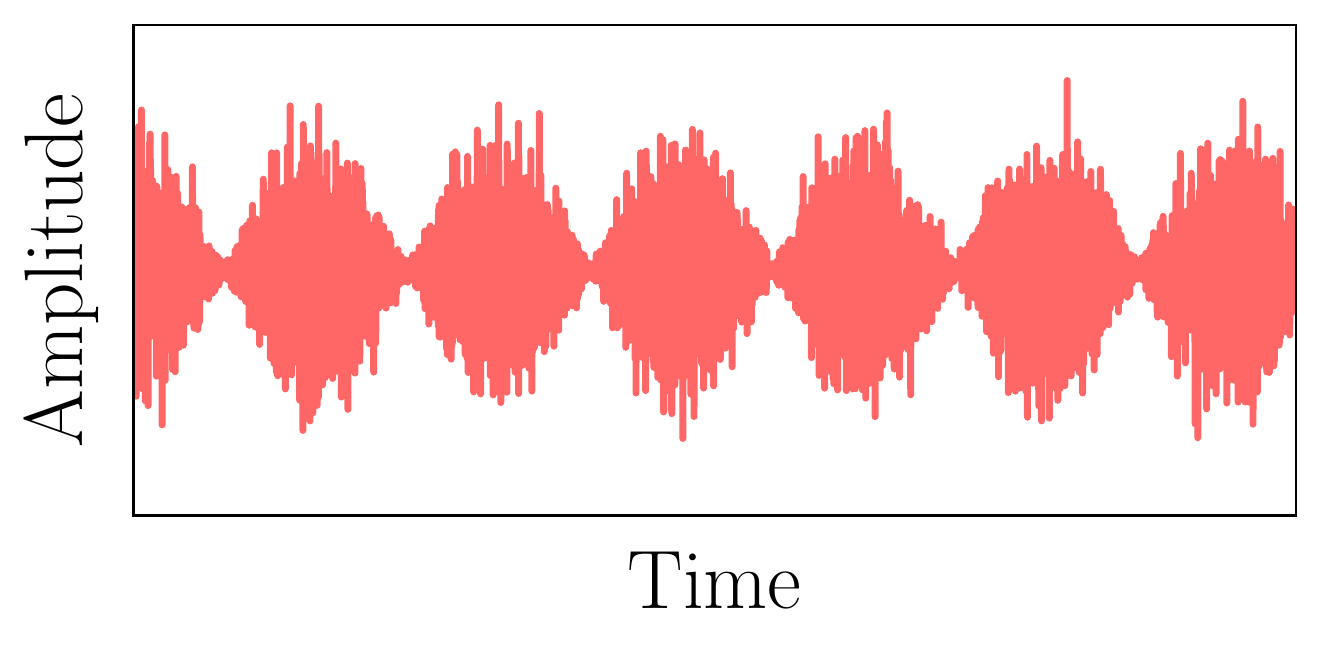}
		\hfill 
		\includegraphics[width=0.24\textwidth, trim={0.2cm 0.2cm 0 0.2cm}, clip]{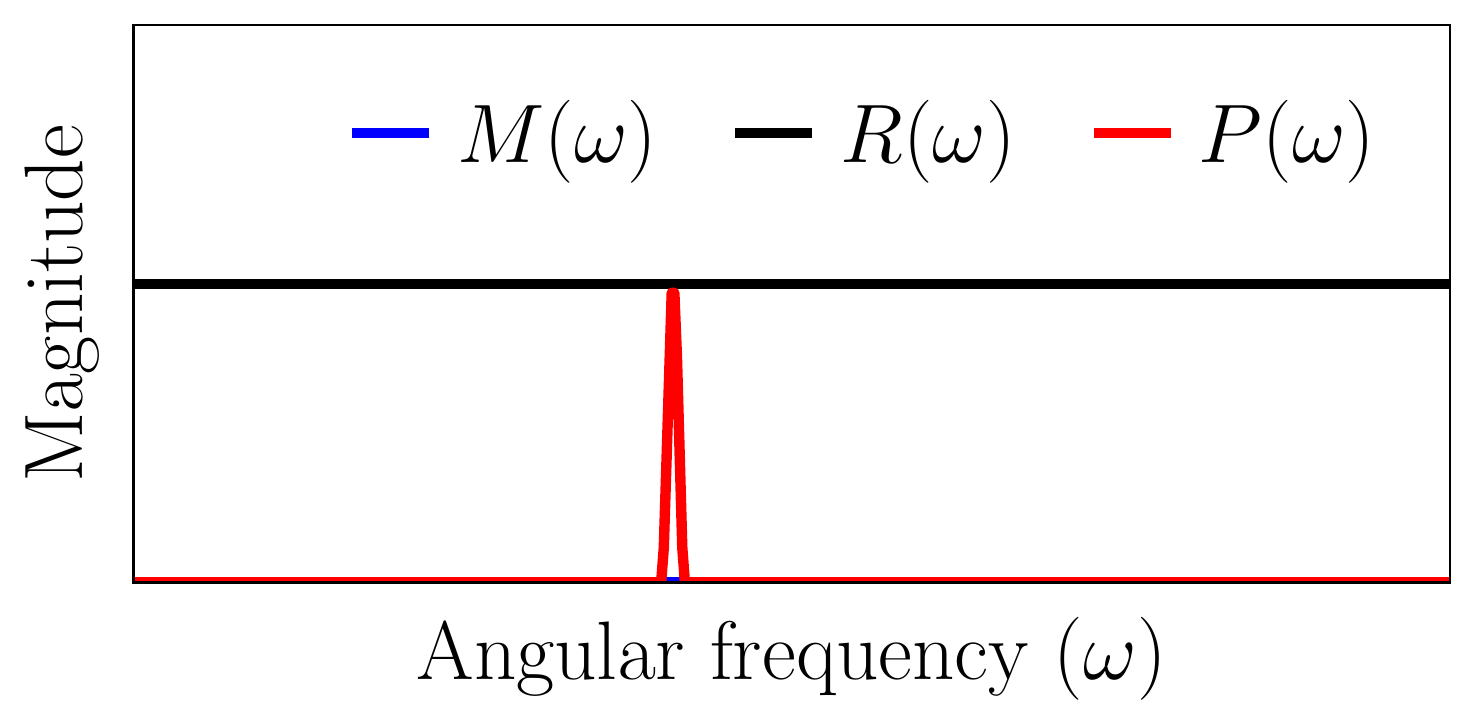} 
		\caption{\small A single realisation of a wide-sense stationary signal (left panel) and its associated spectral moments (right panel).}
		\label{fig:pure_cyclo}
	\end{figure}
	
\end{example}


\begin{example}
	With reference to Remark \ref{remark:PSD}, we next demonstrate the benefits of employing the proposed centred spectral moments over the legacy absolute second-order spectral moments (power spectrum and complementary spectrum). Consider a single realisation of a univariate general nonstationary signal in Fig. \ref{fig:total_signal}. The signal consists of two harmonics at different angular frequencies embedded in general cyclostationary noise (shown in Fig. \ref{fig:constituents}). Observe that the signal constituents are completely \textit{identifiable} when employing the centred spectral moments in Fig. \ref{fig:M_R_P}, whereby: (i) $M(\omega)$ designates the harmonics; (ii) $R(\omega)$ designates the WSS component; and (iii) $P(\omega)$ designates the degree of cyclostationarity. In contrast, the absolute second-order moments, $\tilde{R}(\omega)$ and $\tilde{P}(\omega)$, cannot distinguish between the harmonic and stochastic components, as shown in Fig. \ref{fig:PSD_CSD}.
	
	\vspace{-0.2cm}
	
	\begin{figure}[ht]
		\centering
		\begin{subfigure}[t]{0.24\textwidth}
			\centering
			\includegraphics[width=1\textwidth, trim={0.2cm 0.2cm 0 0.2cm}, clip]{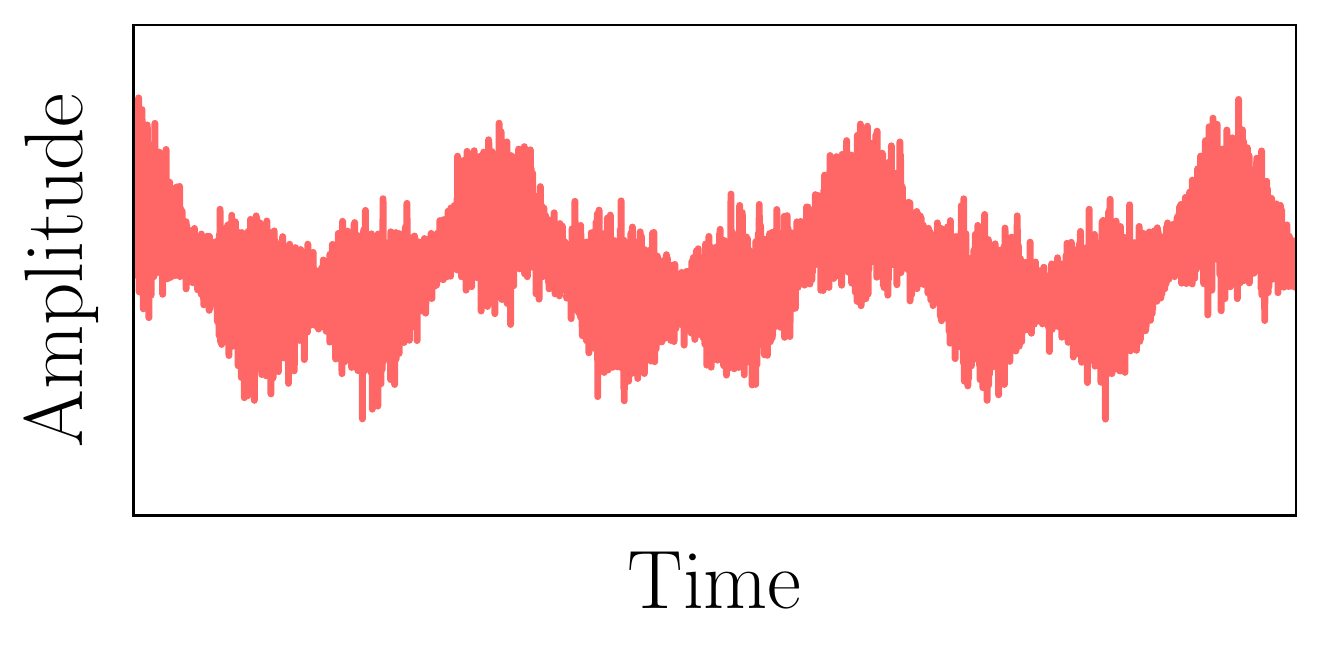} 
			\caption[]%
			{\centering{\small Nonstationary signal.}}    
			\label{fig:total_signal}
		\end{subfigure}
		\hfill
		\begin{subfigure}[t]{0.24\textwidth}
			\centering
			\includegraphics[width=1\textwidth, trim={0.2cm 0.2cm 0 0.2cm}, clip]{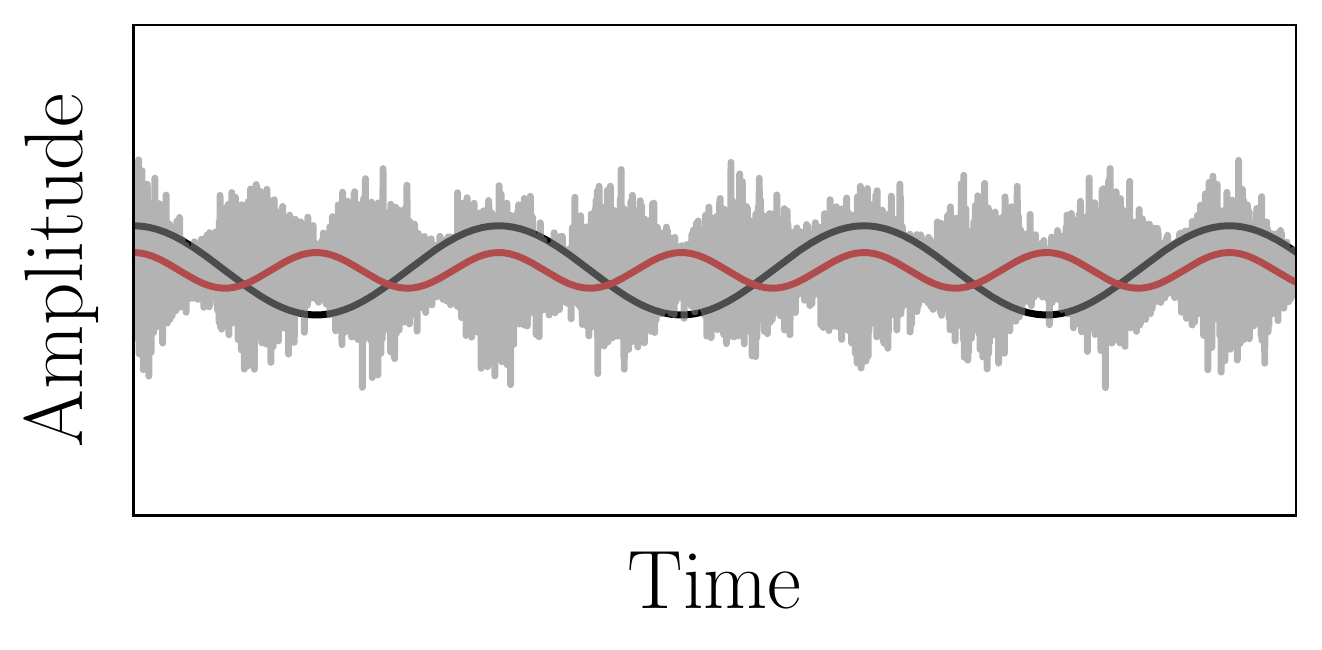} 
			\caption[]%
			{\centering{\small Constituent signals in (a).}}    
			\label{fig:constituents}
		\end{subfigure}
		\vspace{-0.2cm}
		\vskip\baselineskip
		\centering
		\begin{subfigure}[t]{0.24\textwidth}
			\centering
			\includegraphics[width=1\textwidth, trim={0.2cm 0.2cm 0 0.2cm}, clip]{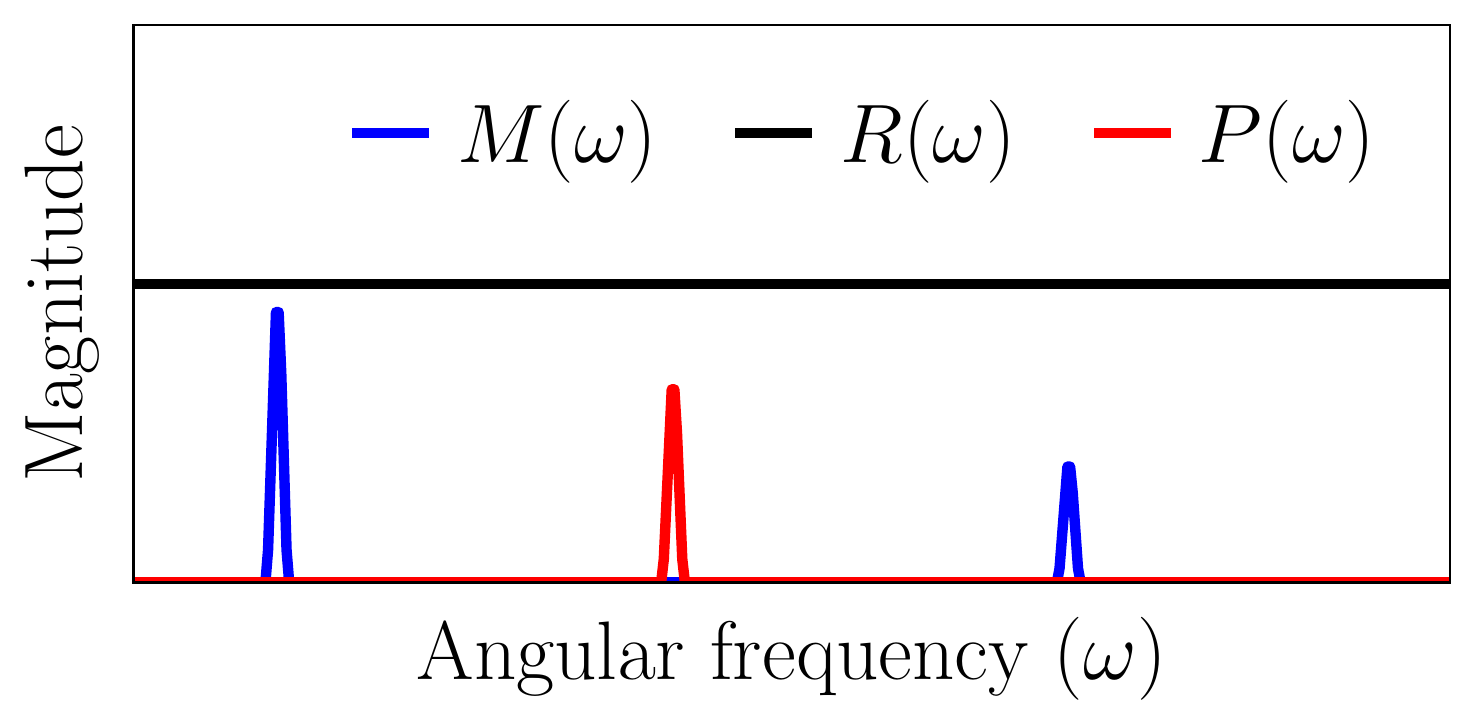} 
			\caption[]%
			{\centering{\small First- and second-order moments of the spectrum.}}    
			\label{fig:M_R_P}
		\end{subfigure}
		\hfill
		\begin{subfigure}[t]{0.24\textwidth}  
			\centering 
			\includegraphics[width=1\textwidth, trim={0.2cm 0.2cm 0 0.2cm}, clip]{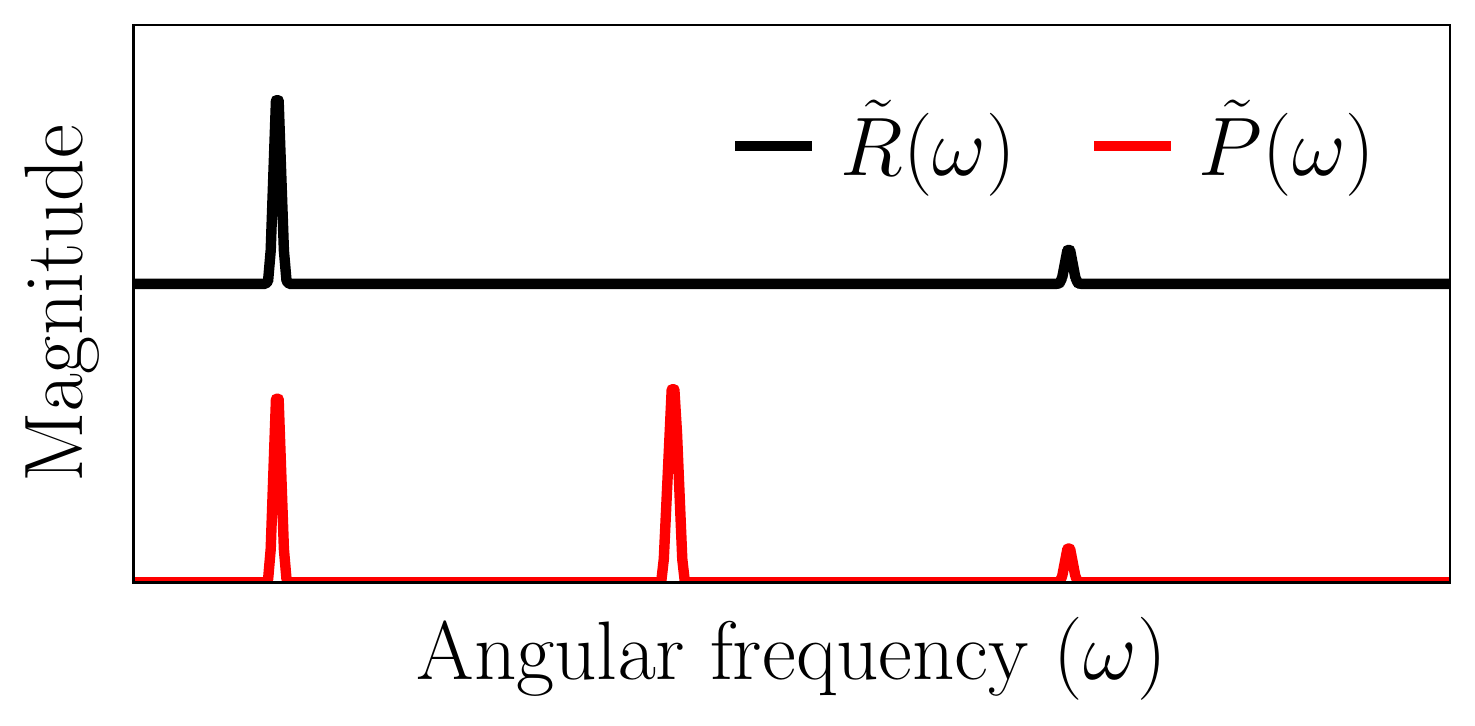} 
			\caption[]%
			{\centering{\small Absolute second-order moments of the spectrum.}}    
			\label{fig:PSD_CSD}
		\end{subfigure}
		\caption[]
		{\small Spectral analysis of a real-valued nonstationary signal. (a) A single realisation. (b) The constituents of the signal in (a). (c) The centred spectral moments. (d) The absolute spectral moments.} 
		\label{fig:nonstationary_example}
	\end{figure}

\end{example}



\section{Augmented spectral statistics}

\label{sec:augmented_spectrum}


Consider a multivariate nonstationary signal which exhibits a discrete frequency spectrum, consisting of $M$ frequency bins, $\boldomega = [\omega_{1},...,\omega_{M}]$. The spectral expansion of $\x(t)$ in (\ref{eq:TFR_expansion_positive}) would therefore become
\begin{equation}
	\x(t) = \frac{1}{\sqrt{2M}} \sum_{m=1}^{M} \!\! \left[ e^{\jmath \omega_{m} t}\bbx(t,\omega_{m}) + e^{-\jmath \omega_{m} t}\bbx^{\ast}(t,\omega_{m}) \right] \label{eq:TFR_expansion_discrete}
\end{equation}

\vspace{-0.2cm}

\begin{remark}
	Unlike the conventional DFT, the normalization by the constant, $\tfrac{1}{\sqrt{2M}}$ in (\ref{eq:TFR_expansion_discrete}), provides a rigorous mapping of coordinates from the time-domain to the time-frequency domain through a pure rotation in the complex plane, thus preserving both the orthogonality and the norm \cite{Smith2007}. 
\end{remark}

\vspace{-0.2cm}

To facilitate the analysis in this work, we express (\ref{eq:TFR_expansion_discrete}) in a compact form by stacking the TFRs associated with each frequency bin to form the matrix
\begin{equation}
	\bbX(t,\boldomega) = \left[ \begin{array}{ccc}
		\bbx(t,\omega_{1}), & \dots, & \bbx(t,\omega_{M})
	\end{array} \right] \in \domC^{N \times M}
\end{equation}
The expansion in (\ref{eq:TFR_expansion_discrete}) can thus be expressed in a matrix form
\begin{equation}
	\x(t) = \frac{1}{\sqrt{2M}} \left( \bbX(t,\boldomega) e^{\jmath \boldOmega t}\1_{M} + \bbX^{\ast}(t,\boldomega) e^{-\jmath \boldOmega t}\1_{M} \right)  \label{eq:derivation_compact_1}
\end{equation}
where $\boldOmega = \diag{\boldomega} \in \domR^{M \times M}$ is the diagonal matrix of angular frequencies, and $\1_{M} \in \domR^{M}$ is a vector of ones. 

Moreover, we can employ the vec-operator and Kronecker product, which together yield $\vect{\A\B} = \left( \B^{\Trans} \otimes \I \right)\vect{\A}$, so as to obtain the following simplified expression
\begin{align}
	\frac{1}{\sqrt{2M}} \bbX(t,\boldomega) e^{\jmath \boldOmega t}\1_{M} & = \frac{1}{\sqrt{2M}}  \left(\1_{M}^{\Trans}e^{\jmath \boldOmega t} \otimes \I_{N}\right) \vect{\bbX(t,\boldomega) }  \notag\\
	& = \boldPhi(t,\boldomega)\bbx(t,\boldomega)
\end{align}
where $\boldPhi(t,\boldomega) \in \domC^{N \times MN}$ is the \textit{spectral basis}, defined as
\begin{align}
	\boldPhi(t,\boldomega) & = \frac{1}{\sqrt{2M}} \left[ 
	\arraycolsep=1pt\def\arraystretch{1.2}
	\begin{array}{ccc}
		e^{\jmath \omega_{1} t}\I_{N}, & \dots, & e^{\jmath \omega_{M} t}\I_{N} \end{array} \right]
\end{align}
with $\I_{N} \in \domR^{N\times N}$ being the identity matrix, and $\bbx(t,\boldomega) \in \domC^{MN}$ the \textit{time-spectrum representation}, given by
\begin{align}
	\bbx(t,\boldomega) = \left[ 
	\def\arraystretch{0.9}
	\begin{array}{c}
		\bbx(t,\omega_{1})\\
		\vdots\\
		\bbx(t,\omega_{M})
	\end{array} \right] \label{eq:time-spectrum_representation}
\end{align}
In this way, we can write (\ref{eq:derivation_compact_1}) as
\begin{equation}
	\x(t) = \boldPhi(t,\boldomega)\bbx(t,\boldomega) + \boldPhi^{\ast}(t,\boldomega)\bbx^{\ast}(t,\boldomega)
\end{equation}
By defining the augmented forms of $\boldPhi(t,\boldomega)$ and $\bbx(t,\boldomega)$ as 
\begin{align}
	\uboldPhi(t,\boldomega) & \! = \! \left[ \! \arraycolsep=1pt\def\arraystretch{1.2} \begin{array}{cc} \boldPhi(t,\boldomega), & \boldPhi^{\ast}(t,\boldomega) \end{array} \! \right] , \!\!\! \quad \ubbx(t,\boldomega) = \left[ \!\! \begin{array}{c}
		\bbx(t,\boldomega)\\
		\bbx^{\ast}(t,\boldomega)
	\end{array} \!\! \right]
\end{align}
we obtain the compact formulation of the spectral expansion in (\ref{eq:TFR_expansion_discrete}), given by
\begin{equation}
	\x(t) = \uboldPhi(t,\boldomega)\ubbx(t,\boldomega) \label{eq:DFT_compact}
\end{equation}
Notice that $\uboldPhi(t,\boldomega)$ is an orthogonal matrix owing to the normalization constant, $\tfrac{1}{\sqrt{2M}}$ in (\ref{eq:TFR_expansion_discrete}), i.e. $\uboldPhi(t,\boldomega)\uboldPhi^{\Her}(t,\boldomega) = \I_{N}$.


\subsection{Augmented spectral statistics}

With the augmented time-spectrum representation, $\ubbx(t,\boldomega) \in \domC^{2MN}$ in (\ref{eq:time-spectrum_representation}), it is now possible to jointly consider all of the dual-frequency spectral covariances in $\boldomega$ through the proposed compact formulation. To see this, consider the following probabilistic model
\begin{equation}
	\ubbx(t,\boldomega) \sim \CNormal{ \ubbm(\boldomega), \ubbR(\boldomega) } \label{eq:probabilistic_spectral_model}
\end{equation}
where $\ubbm(\boldomega) \in \domC^{2MN}$ denotes the \textit{augmented spectral mean}, defined as
\begin{equation}
	\ubbm(\boldomega) = \left[\begin{array}{c}
		\bbm(\boldomega)\\
		\bbm^{\ast}(\boldomega)
	\end{array}\right], \quad \bbm(\boldomega) = \left[
	\def\arraystretch{0.9}
	\begin{array}{c}
		\bbm(\omega_{1})\\
		\vdots\\
		\bbm(\omega_{N})
	\end{array}\right]
\end{equation}
and $\ubbR(\boldomega) \in \domC^{2MN \times 2MN}$ denotes the \textit{augmented spectral covariance}, given by
\begin{align}
	\ubbR(\boldomega) & = \left[\begin{array}{cc}
		\bbR(\boldomega) & \bbP(\boldomega)\\
		\bbP^{\ast}(\boldomega) & \bbR^{\ast}(\boldomega)
	\end{array}\right] \label{eq:augmented_spectral_covariance} \\
	\bbR(\boldomega) & = \left[
	\def\arraystretch{0.9}
	\begin{array}{ccc}
		\bbR(\omega_{1}) & \cdots & \bbR(\omega_{1},\omega_{M}) \\
		\vdots & \ddots & \vdots \\
		\bbR(\omega_{M},\omega_{1}) & \cdots & \bbR(\omega_{M}) 
	\end{array}\right] \label{eq:spectrum_covariance} \\
	\bbP(\boldomega) & = \left[
	\def\arraystretch{0.9}
	\begin{array}{ccc}
		\bbP(\omega_{1}) & \cdots & \bbP(\omega_{1},\omega_{M}) \\
		\vdots & \ddots & \vdots \\
		\bbP(\omega_{M},\omega_{1}) & \cdots & \bbP(\omega_{M})
	\end{array}\right] \label{eq:spectrum_pseudo-covariance}
\end{align}


\noindent By combining the Hermitian and complementary spectral covariances within the augmented form in (\ref{eq:augmented_spectral_covariance}), we obtain the following results which stem from the properties of complex augmented covariance matrices \cite{vandenBos1995,Picinbono1996,Schreier2010}:
\begin{enumerate}[label=\roman*),leftmargin=5mm]
	\item $\bbR(\boldomega)$ is Hermitian symmetric, $\bbR(\boldomega)=\bbR^{\Her}(\boldomega)$, and positive semi-definite, $\bbR(\boldomega) \succcurlyeq 0$;
	\item $\bbP(\boldomega)$ is complex symmetric, $\bbP(\boldomega)=\bbP^{\Trans}(\boldomega)$ ;
	\item The Schur complement of $\ubbR(\boldomega)$ is positive semi-definite, meaning that $\bbR(\boldomega) - \bbP(\boldomega)\bbR^{-\ast}(\boldomega)\bbP^{\ast}(\boldomega) \succcurlyeq 0$;
	\item $\bbR(\boldomega)$ provides an upper bound on the power of $\bbP(\boldomega)$, that is, $\|\bbP(\boldomega)\|_{2} \leq \|\bbR(\boldomega)\|_{2}$.
\end{enumerate}

\begin{remark}
	Observe from (\ref{eq:spectrum_covariance})-(\ref{eq:spectrum_pseudo-covariance}) that all of the dual-frequency spectral covariances are accounted for by the probabilistic spectral model in (\ref{eq:probabilistic_spectral_model}).
\end{remark}

\vspace{-0.5cm}

\subsection{Time-varying statistics in the time-domain}

For convenience, and without loss of generality, from now on we shall drop the frequency spectrum indexing, that is, we shall assume throughout that $\uboldPhi(t,\boldomega) \equiv \uboldPhi(t)$,  $\ubbx(t,\boldomega) \equiv \ubbx(t)$, $\ubbm(\boldomega)\equiv \ubbm$ and $\ubbR(\boldomega)\equiv \ubbR$. In this way, we can write
\begin{equation}
	\x(t) = \uboldPhi(t)\ubbx(t) \label{eq:compact_time_defition}
\end{equation}
The time-domain counterpart, $\x(t)$, has already been shown in (\ref{eq:pdf_time-varying_Gaussian}) to be distributed according to $\x(t) \sim \Normal{\m(t),\R(t)}$, and as a result it exhibits a time-varying pdf of the form
\begin{equation}
	p(\x,t) = \frac{ \exp\left[ -\frac{1}{2} \left(\x(t) - \m(t)\right)^{\Trans}\R^{-1}(t)\left(\x(t) - \m(t)\right) \right] }{\left( 2 \pi \right)^{\frac{N}{2}} \, \det^{\frac{1}{2}}\left( \R (t)\right) } \label{eq:pdf_time_varying}
\end{equation}
It therefore follows that the time-varying pdf can be written in terms of the introduced time-invariant spectral statistics. To see this, we can employ the relation in (\ref{eq:compact_time_defition}) to express the first- and second-order time-varying statistics as follows
\begin{equation}
	\m(t) = \expect{\x(t)}  = \uboldPhi(t)\expect{\ubbx(t)} = \uboldPhi(t)\ubbm \label{eq:time_varying_mean}
\end{equation}
\begin{equation}
\!\!	\R(t) = \cov{\x(t)}  = \uboldPhi(t)\cov{\ubbx(t)}\uboldPhi^{\Her}(t) = \uboldPhi(t)\ubbR \, \uboldPhi^{\Her}(t) \label{eq:time_varying_cov}
\end{equation}
This re-parametrization offers various benefits, as demonstrated in the sequel.

\vspace{-0.2cm}

\subsection{Sampling procedure for a time-varying process}

\label{sec:sampling_procedure}

Now that we have established a probabilistic spectral estimation model, we next derive a sampling procedure for drawing samples, $\x(t) \in \domR^{N}$, from the nonstationary distribution, $\x(t) \sim \Normal{ \m(t), \R(t) }$ in (\ref{eq:pdf_time_varying}). This is achieved by drawing a TFR sample, $\bbx(t) \in \domC^{MN}$, from the stationary TFR distribution, $\ubbx(t) \sim \CNormal{ \ubbm, \ubbR }$. Then, the TFR sample can be transformed into its time-domain counterpart through the Fourier transform in (\ref{eq:compact_time_defition}).

The sampling procedure begins by generating a circularly distributed white Gaussian sample, $\bbw(t) \in \mathbb{C}^{N}$, from a proper multivariate Gaussian distribution $\bbw(t) \sim \CNormal{ \0, \I_{MN} }$. Then, its augmented form, $\ubbw(t) = \left[ \bbw^{\Trans}(t), \bbw^{\Her}(t) \right]^{\Trans}$ is employed to compute the TFR sample
\begin{equation}
	\ubbx(t) = \ubbm + \ubbR^{\frac{1}{2}} \, \ubbw(t)
\end{equation}
where $\ubbR^{\frac{1}{2}}$ is the widely linear Cholesky factor of $\ubbR$. The time-domain signal is obtained through $\x(t) = \uboldPhi(t)\ubbx(t)$ in (\ref{eq:compact_time_defition}).
 

\begin{remark}
	The proposed sampling procedure is closely related to the iterative amplitude adjusted Fourier transform (IAAFT) algorithm \cite{Schreiber1996,Schreiber2000,Mandic2004_2,Mandic2008_2}. Note that the IAAFT algorithm employs the power spectrum and thereby implicitly assumes that the TFR samples are distributed according to
	\begin{equation}
		\ubbx(t) \sim \CNormal{ \0, \diag{\ubbm\,\ubbm^{\Her}+\ubbR} }
	\end{equation}
	This is a consequence of Remark \ref{remark:PSD}. As a result, IAAFT algorithms can only generate WSS samples, while the proposed procedure is capable of generating deterministic and cyclostationary surrogates within a rigorously motivated framework.
\end{remark}


\subsection{Canonical time-frequency coordinates}

The spectral expansion in (\ref{eq:TFR_expansion}) and its compact formulation in (\ref{eq:compact_time_defition}), which our work builds upon, are versions of the \textit{Cram\'er-Lo\`eve} spectral expansion based on the \textit{Fourier basis}. Notice that the TFR, $\ubbs(t) \in \domC^{MN}$, is in general noncircularly distributed and exhibits non-orthogonal frequency increments (see (\ref{eq:dual_frequency_cov})-(\ref{eq:dual_frequency_pcov})), which give rise to several \textit{spectral redundancies}. As such, it may be desirable, or even necessary, to describe instead the time-domain process, $\s(t)$, in terms of circularly distributed TFRs which exhibit orthogonal increments, thereby reducing spectral redundancies. This can be achieved through a \textit{Karhunen-Lo\`eve}-like spectral expansion which employs a basis which is different from the standard Fourier basis.

We therefore consider the task of mapping the noncircular TFR, $\bbs(t)$, to its circular counterpart, denoted by $\bbc(t) \in \domC^{MN}$, using a strictly linear transform, $\boldPsi \in \domC^{MN \times MN}$, that is,
\begin{equation}
	\bbc(t) = \boldPsi \, \bbs(t)
\end{equation}
whereby the second-order statistics of $\bbc$ are given by
\begin{align}
	\!\! \expect{\bbc(t)\bbc^{\Her}(t)} & \! = \! \boldPsi\expect{\bbs(t)\bbs^{\Her}(t)}\boldPsi^{\Her} \! = \! \boldPsi\bbR\boldPsi^{\Her} \! = \! \I \label{eq:SUT_R} \\
	\!\! \expect{\bbc(t)\bbc^{\Trans}(t)} & \! = \! \boldPsi\expect{\bbs(t)\bbs^{\Trans}(t)}\boldPsi^{\Trans} \! = \! \boldPsi\bbP\boldPsi^{\Trans} \! = \! \bbK \label{eq:SUT_P}
\end{align}
where $\bbK \in \domR^{MN \times MN}$ is a diagonal matrix. The diagonal form of the covariances of $\bbc(t)$ indicates that the TFR exhibits orthogonal bin-to-bin increments, that is
\begin{equation}
	\!\! \expect{\bbc(t,\omega)\bbc^{\Her}(t,\nu)}=\expect{\bbc(t,\omega)\bbc^{\Trans}(t, \nu)}=\0, \quad \omega \neq \nu
\end{equation}
The TFRs which satisfy these conditions are said to be \textit{strongly uncorrelated}. In this way, we can then express the time-domain process through the following spectral expansion
\begin{align}
	\s(t) & = \boldPhi(t)\bbs(t) + \boldPhi^{\ast}(t)\bbs^{\ast}(t) \notag\\
	& = \boldPhi(t)\boldPsi^{-1}\,\bbc(t) + \boldPhi^{\ast}(t)\boldPsi^{-\ast}\,\bbc^{\ast}(t)
\end{align}
where $\boldPhi(t)\boldPsi^{-1}$ is the spectral basis which maps $\s(t)$ to circularity distributed TFRs. Notice that, because the transform is strictly linear, the Hermitian symmetry of the spectrum of $\s(t)$ is preserved in this formulation.

It then follows that the transform, $\boldPsi$, can be obtained through the \textit{strong uncorrelating transform} (SUT) \cite{DeLathauwer2002,Eriksson2004,Mandic2012_2,Mandic2015_4}, which performs the simultaneous diagonalization of $\bbR$ and $\bbP$ in (\ref{eq:SUT_R})-(\ref{eq:SUT_P}), which we elaborate upon next. 

We begin by defining the \textit{spectral coherence} matrix as the pre-whitened version of $\bbP$, defined as
\begin{equation}
	\bbC = \bbR^{-\frac{1}{2}}\bbP\bbR^{-\frac{\Trans}{2}}
\end{equation}
Since $\bbC$ is complex symmetric, we have $\bbC = \bbC^{\Trans}$, and not Hermitian symmetric, so that $\bbC \neq \bbC^{\Her}$, there exists a special singular value decomposition for complex symmetric matrices -- the so called \textit{Takagi factorization} -- which yields
\begin{equation}
	\bbC = \V\bbK\V^{\Trans}
\end{equation}
where $\V \in \domC^{MN \times MN}$ is unitary, $\V^{\Her}\V = \V\V^{\Her} = \I$, and $\bbK = \diag{\kappa_{1},\dots,\kappa_{NM}}$ is a diagonal matrix containing the real-valued canonical correlations, which we refer to as \textit{spectral circularity coefficients}, $1\geq \kappa_{1} \geq \cdots \geq \kappa_{NM} \geq 0$. 

By defining a new transform as 
\begin{equation}
	\boldPsi = \V^{\Her}\bbR^{-\frac{1}{2}}
\end{equation}
we obtain the simultaneous diagonalization of $\bbR$ and $\bbP$ in (\ref{eq:SUT_R})-(\ref{eq:SUT_P}), since
\begin{align}
	\boldPsi\bbR\boldPsi^{\Her} & =  \V^{\Her}\bbR^{-\frac{1}{2}}\bbR\bbR^{-\frac{\Her}{2}}\V = \V^{\Her}\V = \I \\
	\boldPsi\bbP\boldPsi^{\Trans} & =  \V^{\Her}\bbR^{-\frac{1}{2}}\bbP\bbR^{-\frac{\Trans}{2}}\V^{\ast} = \V^{\Her}\bbC\V =  \bbK 
\end{align}
The proposed spectral description then becomes
\begin{equation}
	\bbc(t) = \V^{\Her}\bbR^{-\frac{1}{2}}\bbs(t)
\end{equation}
and is said to be given in \textit{canonical coordinates} \cite{Eriksson2006}. Upon defining the augmented versions of $\boldPsi$ and $\bbc(t)$ as
\begin{equation}
	\uboldPsi = \left[\begin{array}{cc}
		\boldPsi&\0\\
		\0 & \boldPsi^{\ast}
	\end{array}\right], \quad \ubbc(t) = \left[\begin{array}{c}
		\bbc(t)\\
		\bbc^{\ast}(t)
	\end{array}\right]
\end{equation}
we can obtain a compact formulation for the proposed spectral expansion in terms of canonical coordinates, given by
\begin{equation}
	\s(t) = \uboldPhi(t)\uboldPsi^{-1}\ubbc(t) \label{eq:SUT_basis}
\end{equation}

\begin{remark}
	The change of basis provided by (\ref{eq:SUT_basis}) is particularly useful for spectral analysis purposes because it reduces the number of distinct spectral covariance parameters required to describe the process from $(N^{2}M^{2}+NM)$ in $\bbR$ and $\bbP$ to $NM$ in $\bbK$. 
\end{remark}

\begin{remark}
	The $i$-th spectral circularity coefficient, $\kappa_{i}$, quantifies the degree of circularity exhibited by the $i$-th entry of the TFR, $\bbc(t) \in \domC^{NM}$. In the time-domain, this is equivalent to quantifying the degree of cyclostationarity exhibited by $\s(t)$ associated with the $i$-th element in $\bbc(t)$. Therefore, if there is at least one non-zero spectral circularity coefficient, $\kappa_{i}>0$, then $\s(t)$ will exhibit a cyclostationary behaviour. Otherwise, if all coefficients are equal to zero, then $\kappa_{i} = 0$ for all $i$, and $\s(t)$ will be WSS.
\end{remark}

\begin{remark}
	The SUT has also been employed in \cite{Okopal2014,Okopal2015}, whereby the spectral circularity coefficients were used to identify time–frequency regions that contain voiced speech. However, we note that the work in \cite{Okopal2014,Okopal2015} performed the SUT on the absolute second-order spectral moments, $\tilde{\bbR}(\omega)$ and $\tilde{\bbP}(\omega)$ in (\ref{eq:PSD})-(\ref{eq:CSD}), meaning that both harmonics and cyclostationary behaviour were detected by the spectral circularity measure. In turn, our approach performs the SUT on centred moments, thereby targetting the cyclostationary information only.
\end{remark}


\section{Statistically Consistent Spectral Estimation}

\label{sec:MLE}

\subsection{Maximum likelihood (ML) estimation}

The log-likelihood of $T$ samples, $\x(t) \in \domR^{N}$, which are distributed according to the nonstationary distribution in (\ref{eq:pdf_time_varying}), and parametrized by the parameters $\ubbm$ and $\ubbR$, is given by
\begin{align}
	\mathcal{L} & = \ln\left( \prod_{t=0}^{T-1} p(\x,t) \right) = \sum_{t=0}^{T-1} \ln\left( p(\x,t) \right) \\
	& = - \frac{1}{2} \biggl( TN \ln(2\pi) + \sum_{t=0}^{T-1} \ln \det (\R(t)) +  \s^{\Trans}(t) \R^{-1}(t) \s(t) \biggr) \notag
\end{align}
where $\R(t) = \uboldPhi(t)\ubbR \, \uboldPhi^{\Her}(t)$ and $\s(t) = \x(t) - \uboldPhi(t)\ubbm$. 

To determine the maximum likelihood estimator of $\ubbm$, we can evaluate $\frac{\partial \mathcal{L}}{\partial \ubbm}=\0$ to yield the stationary point
\begin{align}
\frac{\partial \mathcal{L}}{\partial \ubbm} & = -2 \! \sum_{t=0}^{T-1} \uboldPhi^{\Her}(t) \! \left( \uboldPhi(t) \ubbR \uboldPhi^{\Her}(t)\right)^{\! -1} \!\! \left( \x(t) - \uboldPhi(t)\ubbm \right) = \0
\end{align}
which, after rearranging, leads to the ML estimator of the spectral mean, $\ubbm$, given by
\begin{equation}
	\hat{\ubbm} \! = \! \left(  \sum_{t=0}^{T-1} \uboldPhi^{\Her}(t)\R^{-1}\!(t) \uboldPhi(t) \! \right)^{\!\!\!\!-1} \!\!\!\! \left(  \sum_{t=0}^{T-1} \uboldPhi^{\Her}(t)\R^{-1}\!(t) \x(t) \! \right) \label{eq:ML_mean}
\end{equation}

\begin{remark} \label{remark:rank_deficient_basis}
	The term, $\R^{-1}(t) \! = \! \left( \uboldPhi(t)\ubbR \, \uboldPhi^{\Her}(t)  \right)^{\! -1}$ in (\ref{eq:ML_mean}), cannot be simplified because $\uboldPhi(t)$ is rank deficient. Therefore, $\uboldPhi^{\Her}(t)$ is not injective and there exists a vector $\v \in \domC^{2MN} \backslash \{\0\}$ which achieves $\uboldPhi^{\Her}(t)\v = \0$ and hence $\uboldPhi(t)\ubbR \, \uboldPhi^{\Her}(t)\v = 0$. It thus follows that $\R(t) = \uboldPhi(t)\ubbR \, \uboldPhi^{\Her}(t)$ is not injective and thus not invertible.
\end{remark}

\begin{remark} \label{remark:ML_time_varying_mean}
	The ML estimator of the time-varying mean, $\m(t)$, is obtained directly through the relationship in (\ref{eq:time_varying_mean}), i.e.
	\begin{align}
		\hat{\m}(t) = \uboldPhi(t)\hat{\ubbm} \label{eq:ML_time_varying_mean}
	\end{align}
\end{remark}

To determine the ML estimator of the spectral covariance, $\ubbR$, we evaluate the stationary point, $\frac{\partial \mathcal{L}}{\partial \ubbR}=\0$, to yield
\begin{equation}
\frac{\partial \mathcal{L}}{\partial \ubbR} = \sum_{t=0}^{T-1} \ubbR^{-1} \! - \uboldPhi^{\Her}(t)\R^{-1}(t) \s(t)\s^{\Trans}(t) \R^{-1}(t)\uboldPhi(t) = \0
\end{equation} 
Upon computing the centred time-domain signal using the estimate of $\m(t)$ in (\ref{eq:ML_time_varying_mean}), $\s(t) = \x(t) - \hat{\m}(t)$, we obtain the ML estimator of $\ubbR$ in the form
\begin{equation}
	\ubbR = \left( \frac{1}{T} \sum_{t=0}^{T-1} \uboldPhi^{\Her}(t)\R^{-1}(t) \s(t)\s^{\Trans}(t) \R^{-1}(t)\uboldPhi(t) \right)^{-1} \label{eq:ML_cov}
\end{equation}

\begin{remark}
	Notice that the maximum likelihood estimate of $\ubbR$ in (\ref{eq:ML_cov}) is a fixed-point solution, since it is a function of the term $\R^{-1}(t) = \left( \uboldPhi(t) \ubbR \, \uboldPhi^{\Her}(t) \right)^{-1}$, which is dependent of $\ubbR$.
\end{remark}

\begin{remark}
	Following on from Remark \ref{remark:ML_time_varying_mean}, we can directly obtain the ML estimator of the time-varying covariance, $\R(t)$, through the Fourier transform relationship in (\ref{eq:time_varying_cov}), that is
	\begin{equation}
		\hat{\R}(t) = \uboldPhi(t)\hat{\ubbR}\,\uboldPhi^{\Her}(t) \label{eq:ML_time_varying_cov}
	\end{equation}
\end{remark}

\subsection{Approximative solution}

Several issues may arise when performing the maximum likelihood procedure in practice. For instance, the procedure is iterative and therefore susceptible to local maxima, since the estimate of $\ubbm$ in (\ref{eq:ML_mean}) requires an estimate of $\ubbR$ in (\ref{eq:ML_cov}) and vice versa. Furthermore, each estimate involves multiple matrix inversions which may become computationally intractable for large dimensions, $N$ and $M$, or may even not exist for small data lengths, $T$.

To this end, we consider the following approximative maximum likelihood estimation procedure, which instead employs the likelihood function of the the time-spectrum representation, $\ubbx(t) \in \domC^{2MN}$, given by
\begin{equation}
	p(\ubbx,t) = \frac{ \exp \left[  - \frac{1}{2}  \left( \ubbx(t) -  \ubbm \right)^{\Her} \ubbR^{-1}  \left( \ubbx(t) -  \ubbm \right) \right] }{\pi^{MN} \det^{\frac{1}{2}} ( \ubbR )} \label{eq:pdf_time_spectrum}
\end{equation}
By replacing $\ubbx(t)$ in (\ref{eq:pdf_time_spectrum}) with its least squares estimate based on (\ref{eq:compact_time_defition}), i.e.
\begin{align}
	\hat{\ubbx}(t) = \uboldPhi^{+}(t)\x(t) \equiv \uboldPhi^{\Her}(t)\x(t)
\end{align}
with the symbol $(\cdot)^{+}$ denoting the pseudo-inverse operator, the likelihood of $\ubbx(t)$ can now be expressed in terms of the observable, $\x(t)$. We therefore obtain
\begin{equation}
p(\x,t) \simeq  \frac{ \exp \left[  - \frac{1}{2}  \left( \uboldPhi^{\Her}(t)\x(t) -  \ubbm \right)^{\Her} \ubbR^{-1}  \left( \uboldPhi^{\Her}(t)\x(t) -  \ubbm \right) \right] }{\pi^{MN} \det^{\frac{1}{2}} ( \ubbR )} \label{eq:pdf_approx}
\end{equation}
By maximising the log-likelihood of $T$ samples, $\uboldPhi^{\Her}(t)\x(t)$, which are distributed according to the pdf in (\ref{eq:pdf_approx}), given by $\mathcal{L} = \ln\biggl( \prod_{t=0}^{T-1} p(\x,t) \biggr)$, we find that the ML estimators reduce to
\begin{equation}
	\hat{\ubbm} =\frac{1}{T} \sum_{t=0}^{T-1} \uboldPhi^{\Her}(t) \x(t) \label{eq:ML_approx_mean}
\end{equation}
\begin{equation}
\hat{\ubbR} = \frac{1}{T} \sum_{t=0}^{T-1} \uboldPhi^{\Her}(t) \hat{\s}(t)\hat{\s}^{\Trans}(t) \uboldPhi(t) \label{eq:ML_approx_cov}
\end{equation}
with $\hat{\s}(t) = \x(t) - \hat{\m}(t) = \x(t) - \uboldPhi(t)\hat{\ubbm}$.

\begin{remark}
	Recall from Remark \ref{remark:rank_deficient_basis} that the basis matrix, $\uboldPhi(t) \in \domC^{N \times 2MN}$, is rank-deficient. Therefore, there are infinitely many solutions to the linear system $\x(t) = \uboldPhi(t)\ubbx(t)$, since $\uboldPhi^{\Her}(t)$ has a non-empty null space. By employing the least squares solution based on the Moore-Penrose pseudo-inverse, i.e. $\ubbx(t) = \uboldPhi^{\Her}(t)\x(t)$, we obtain the solution to $\ubbx(t)$ which minimizes the Euclidean distance, $\| \x(t) - \uboldPhi(t)\ubbx(t) \|_{2}$. Furthermore, out of all possible solutions which attain the least squares error, this solution also provides the estimate with the smallest norm, $\| \ubbx(t) \|_{2}$.
\end{remark}

\begin{remark}
	The approximate ML estimator of $\ubbm$ in (\ref{eq:ML_approx_mean}) is, in essence, the \textit{discrete Fourier transform} (DFT) of $\x(t)$. In other words, the DFT represents the sample mean of the \textit{sample time-spectrum}. Similarly, the estimator of $\ubbR$ in (\ref{eq:ML_approx_cov}) is the power spectrum matrix of the centred variable, $\s(t)$. 
\end{remark}


A natural next step is to consider whether an estimate of $\bbs(t)$ can be retrieved from $\s(t)$. In general, this is not possible, owing to the indeterminacy of the problem formulation, since for every real-valued observation, $\s(t)$, there are two real-valued unknowns, $\Real{\bbs(t)}$ and $\Imag{\bbs(t)}$. Nonetheless, it is possible to estimate the value of $\bbs(t)$ in the minimum mean-square error sense. Observe that the real and imaginary parts of $\bbs(t)$ are related through the strictly linear mapping
\begin{align}
	\bbs^{\ast}(t) = \W \bbs(t)
\end{align}
where $\W$ is the \textit{canonical correlation matrix} between the TFR and its own complex conjugate form. Therefore, we can estimate $\W$ by minimising the mean-square error
\begin{align}
	\W & = \arg\min_{\W} \; \expect{\| \bbs^{\ast}(t) - \W \bbs(t)  \|_{2}}  = \bbP^{\ast}\bbR^{-1}
\end{align}
to obtain the following relationship
\begin{align}
	\s(t) & = \uboldPhi(t)\ubbs(t) = \left( \boldPhi(t) + \boldPhi^{\ast}(t)\bbP^{\ast}\bbR^{-1} \right)\bbs(t) \label{eq:s_time_to_s_TFR}
\end{align}
The optimal estimate of $\bbs(t)$, in the MSE sense, becomes
\begin{align}
	\bbs(t) = \left( \boldPhi(t) + \boldPhi^{\ast}(t)\bbP^{\ast}\bbR^{-1} \right)^{+}\s(t)
\end{align}

\begin{remark}
	For the univariate case ($N=1$), the basis matrix in (\ref{eq:s_time_to_s_TFR}) reduces to $\left( e^{\jmath \omega t} + \varrho^{\ast}(\omega) e^{-\jmath \omega t} \right)$, which is the \textit{elliptic Fourier basis} considered in \cite{Schreier2008,Walden2011,Walden2012}, whereby the quantity $\varrho(\omega) = P(\omega)/R(\omega)$ is the \textit{spectral circularity coefficient}. The term $\left( \boldPhi(t) + \boldPhi^{\ast}(t)\bbP^{\ast}\bbR^{-1} \right)$ is therefore the multivariate extension of the \textit{elliptic Fourier basis}.
\end{remark}


\subsection{Conditions for statistical consistency}


In practice, a single realisation of $T$ samples of the signal, $\x(t)$, is typically  observed, whereby the conditions of the occurrence cannot be duplicated or repeated. As a result, we cannot employ the expectation (ensemble-average) operator to evaluate the statistics of $\x(t)$. Instead, we must resort to statistical estimates based on the the time-average operator, as in (\ref{eq:ML_approx_mean}) and (\ref{eq:ML_approx_cov}). However, it is not obvious if the time-average based estimate asymptotically approaches its ensemble-average counterpart. If the time-average based estimate is evaluated from a realisation of $T$ samples approaches the corresponding ensemble-average counterpart in the limit, $T \to \infty$, then the estimator is said to be \textit{statistically consistent}.

By virtue of the law of large numbers, the condition of statistical consistency will hold if the spectral samples, $\ubbx(t)$, are independent and identically distributed (\textit{i.i.d.}). However, if the samples are highly correlated then taking more samples would not necessarily lead to a statistically consistent estimate. More specifically, the \textit{spectral autocovariance} and \textit{autoconvolution functions} play an important role in determining the conditions of statistical consistency. 

\subsubsection{Statistical consistency of $\hat{\bbm}$}

From (\ref{eq:ML_approx_mean}) it is easy to see that $\hat{\bbm}$ is an unbiased estimate of the true spectral mean, $\bbm$. However, for the estimate to be statistically consistent it is required that if the variance vanishes, $\hat{\bbm}$ converges to $\bbm$ as $T \to \infty$. This convergence will occur (in the mean square sense) if the \textit{estimate covariance} of $\bbm$, that is, $\cov{\hat{\bbm}}$ vanishes with an increase in the number of samples. To determine the conditions for when this occurs, we begin by evaluating the estimated covariance of $\bbm$ as follows
\begin{align}
\cov{\hat{\bbm}} & = \expect{ \left( \hat{\bbm} - \bbm \right) \left( \hat{\bbm} - \bbm  \right)^{\Her} } \notag\\
& = \expect{ \left( \frac{1}{T} \sum_{t=0}^{T-1} \bbs(t_{n}) \right)\left( \frac{1}{T} \sum_{t=0}^{T-1} \bbs(t) \right)^{\Her} } \notag\\
& = \frac{1}{T^{2}} \sum_{t_{1}=0}^{T-1} \sum_{t_{2}=0}^{T-1}\expect{\bbs(t_{1})\bbs^{\Her}(t_{2})} \notag\\
& = \frac{1}{T^{2}} \sum_{t=0}^{T-1} \sum_{t=0}^{T-1}\bbR(t_{1}-t_{2}) \label{eq:estimate_covariance_m}
\end{align}
where $\bbR(t_{1}-t_{2}) = \expect{\bbs(t_{1})\bbs^{\Her}(t_{2})}$ is the \textit{spectral autocovariance function}, with $\bbR(0) \equiv \bbR$. Upon inserting $\tau = (t_{1}-t_{2})$ into (\ref{eq:estimate_covariance_m}) and converting the double summation into a single summation, we obtain
\begin{equation}
\cov{\hat{\bbm}} = \frac{1}{T^{2}} \sum_{\tau=0}^{T-1} \left( T - |\tau| \right) \bbR(\tau) = \frac{1}{T} \sum_{\tau=0}^{T-1} \left( 1 - \frac{|\tau|}{T} \right) \bbR(\tau)
\end{equation}
Therefore, the estimate, $\hat{\bbm}$, is statistically consistent if the following convergence condition holds
\begin{equation}
\lim_{T \to \infty} \frac{1}{T}\sum_{\tau=0}^{T-1} \left( 1 - \frac{|\tau|}{T} \right) \bbR(\tau) = \0
\end{equation}


\subsubsection{Statistical consistency of $\hat{\bbR}$}

To investigate the condition for statistical consistency of $\hat{\bbR}$, we begin by evaluating the estimated covariance of $\bbR$, given by
\begin{align}
\mathsf{cov}\!\left\{\hat{\bbR}\right\} \! & = \expect{ \left( \hat{\bbR} - \bbR \right) \left( \hat{\bbR} - \bbR  \right)^{\Her} } \notag\\
& = \! \expect{ \!\!\! \left( \! \frac{1}{T} \! \sum_{t=0}^{T-1} \! \bbs(t)\bbs^{\Her}\!(t)  - \bbR \! \right) \!\!\! \left( \! \frac{1}{T} \! \sum_{t=0}^{T-1} \! \bbs(t)\bbs^{\Her}\!(t) - \bbR \! \right)^{\!\!\! \Her} } \notag\\
& = \frac{1}{T^{2}} \sum_{t_{1}=0}^{T-1} \sum_{t_{2}=0}^{T-1}\expect{ \bbs(t_{1})\bbs^{\Her}(t_{1})\bbs(t_{2})\bbs^{\Her}(t_{2})} - \bbR\bbR \label{eq:estimate_covariance_R}
\end{align}
By virtue of Isserlis' theorem \cite{Isserlis1918}, the fourth order moment of $\bbs(t)$ can be expanded as follows
\begin{align}
&\expect{\bbs(t_{1})\bbs^{\Her}(t_{1})\bbs(t_{2})\bbs^{\Her}(t_{2})} \notag\\
& \quad\quad\quad = \expect{\bbs(t_{1})\bbs^{\Her}(t_{1})}\expect{\bbs(t_{2})\bbs^{\Her}(t_{2})} \notag\\
& \quad\quad\quad + \expect{\bbs(t_{1})\bbs^{\Trans}(t_{2})}\expect{\bbs^{\ast}(t_{1})\bbs^{\Her}(t_{2})} \notag\\
& \quad\quad\quad + \expect{\bbs(t_{1})\bbs^{\Her}(t_{2})}\expect{\bbs^{\ast}(t_{1})\bbs^{\Trans}(t_{2})}  \notag\\
& \quad\quad\quad = \bbR\,\bbR + \bbP(t_{1}-t_{2})\bbP^{\ast}(t_{1}\!-\!t_{2}) + \bbR(t_{1}\!-\!t_{2})\bbR^{\ast}(t_{1}\!-\!t_{2})
\end{align}
where $\bbP(t_{1}-t_{2}) = \expect{\bbs(t_{1})\bbs^{\Trans}(t_{2})}$ is the \textit{spectral autoconvolution function}, with $\bbP(0) \equiv \bbP$. Upon employing the variable substitution, $\tau = (t_{1}-t_{2})$, we obtain the condition for the statistically consistency of $\hat{\bbR}$ to hold, given by
\begin{align}
	\lim_{T \to \infty} \frac{1}{T} \sum_{t=0}^{T-1} \left( 1 - \frac{|\tau|}{T} \right)\left( \bbP(\tau)\bbP^{\ast}(\tau) + \bbR(\tau)\bbR^{\ast}(\tau) \right) = \0 \label{eq:statistical_consistency_R}
\end{align}

\subsubsection{Statistical consistency of $\hat{\bbP}$}

To determine the condition for statistical consistency in $\hat{\bbP}$, we can compute the covariance of $\bbP$, given by
\begin{align}
	\mathsf{cov}\!\left\{\hat{\bbP}\right\} \! & = \expect{ \left( \hat{\bbP} - \bbP \right) \left( \hat{\bbP} - \bbP  \right)^{\Her} } \notag\\
	& = \! \expect{ \!\!\! \left( \! \frac{1}{T} \! \sum_{t=0}^{T-1} \! \bbs(t)\bbs^{\Trans}\!(t)  - \bbP \! \right) \!\!\! \left( \! \frac{1}{T} \! \sum_{t=0}^{T-1} \! \bbs(t)\bbs^{\Trans}\!(t) - \bbP \! \right)^{\!\!\! \Her} } \notag\\
	& = \frac{1}{T^{2}} \sum_{t_{1}=0}^{T-1} \sum_{t_{2}=0}^{T-1}\expect{ \bbs(t_{1})\bbs^{\Trans}(t_{1})\bbs^{\ast}(t_{2})\bbs^{\Her}(t_{2})} - \bbP\bbP^{\Her} 
\end{align}
whereby the fourth-order moment admits the expansion
\begin{align}
	& \expect{\bbs(t_{1})\bbs^{\Trans}(t_{1})\bbs^{\ast}(t_{2})\bbs^{\Her}(t_{2})} \notag\\
	& \quad\quad\quad = \expect{\bbs(t_{1})\bbs^{\Trans}(t_{1})}\expect{\bbs^{\ast}(t_{2})\bbs^{\Her}(t_{2})} \notag\\
	& \quad\quad\quad + 2\expect{\bbs(t_{1})\bbs^{\Her}(t_{2})}\expect{\bbs(t_{1})\bbs^{\Her}(t_{2})} \notag\\
	& \quad\quad\quad = \bbP\,\bbP^{\Her} + 2\, \bbR(t_{1}-t_{2})\bbR^{\ast}(t_{1}-t_{2})
\end{align}
Therefore, with the variable substitution, $\tau = (t_{1}-t_{2})$, we can show that the estimate $\hat{\bbP}$ is statistically consistent if the following convergence holds
\begin{align}
	\lim_{T \to \infty} \frac{1}{T} \sum_{t=0}^{T-1} 2 \left( 1 - \frac{|\tau|}{T} \right)\bbR(\tau)\bbR^{\ast}(\tau)  = \0 \label{eq:statistical_consistency_P}
\end{align}


\begin{remark}
	Notice from (\ref{eq:statistical_consistency_R}) and (\ref{eq:statistical_consistency_P}) that the condition for statistical consistency of $\hat{\bbR}$ is milder than that for $\hat{\bbP}$. The rate at which the time-average converges to the ensemble-average is equivalent for $\hat{\bbR}$ and $\hat{\bbP}$ only when the TFRs are maximally noncircular (rectilinear) whereby $\|\bbP(\tau)\|_{2}=\|\bbR(\tau)\|_{2}$. For general non-rectilinear TFRs, whereby $\|\bbP(\tau)\|_{2}<\|\bbR(\tau)\|_{2}$, the estimator of $\hat{\bbR}$ achieves a lower estimation variance that the estimator of $\hat{\bbP}$.
\end{remark}

\begin{remark} \label{remark:statistical_consistency}
	If the TFR samples, $\bbx(t)$, are \textit{i.i.d.}, with $\bbR(\tau)=\bbP(\tau)=\0$, $\tau \neq 0$, the rate at which the time-average based estimates converge to their ensemble-average counterparts reduces to
	\begin{align}
		\cov{\hat{\bbm}} & = \frac{1}{T} \, \bbR \\
		\cov{\hat{\bbR}} & = \frac{1}{T}  \left( \bbP \, \bbP^{\ast} + \bbR \, \bbR \right) \\
		\cov{\hat{\bbP}} & = \frac{1}{T}  \left( 2 \, \bbR \, \bbR \right)
	\end{align}
	and the estimates are therefore asymptotically consistent.
\end{remark}

\begin{example}
	Following on from Remark \ref{remark:statistical_consistency}, the statistical consistency of the proposed spectral estimators was verified empirically. The sampling procedure derived in Section \ref{sec:sampling_procedure} was employed to generate $T$ univariate samples ($N=1$), $x(t) \in \domR$, from a time-varying Gaussian process consisting of one harmonic at an angular frequency $\omega_{0}$ embedded in WSS noise. The parameters of the distribution were chosen arbitrarily. The estimation variance was evaluated for each spectral moment using sample lengths, $T$, in the range $[10,10^{6}]$. The results were averaged over 1000 independent Monte Carlo simulations and are displayed in Fig. \ref{fig:statistical_consistency}. Observe that the attained asymptotic convergence of the estimators demonstrates their \textit{statistical consistency}, and that $\var{\hat{R}(\omega_{0})}<\var{\hat{P}(\omega_{0})}$, since $|P(\omega_{0})|=0$.

	\begin{figure}[h!]
		\centering
		\vspace{-0.4cm}
		\includegraphics[width=0.475\textwidth, trim={0 0 0 0}, clip]{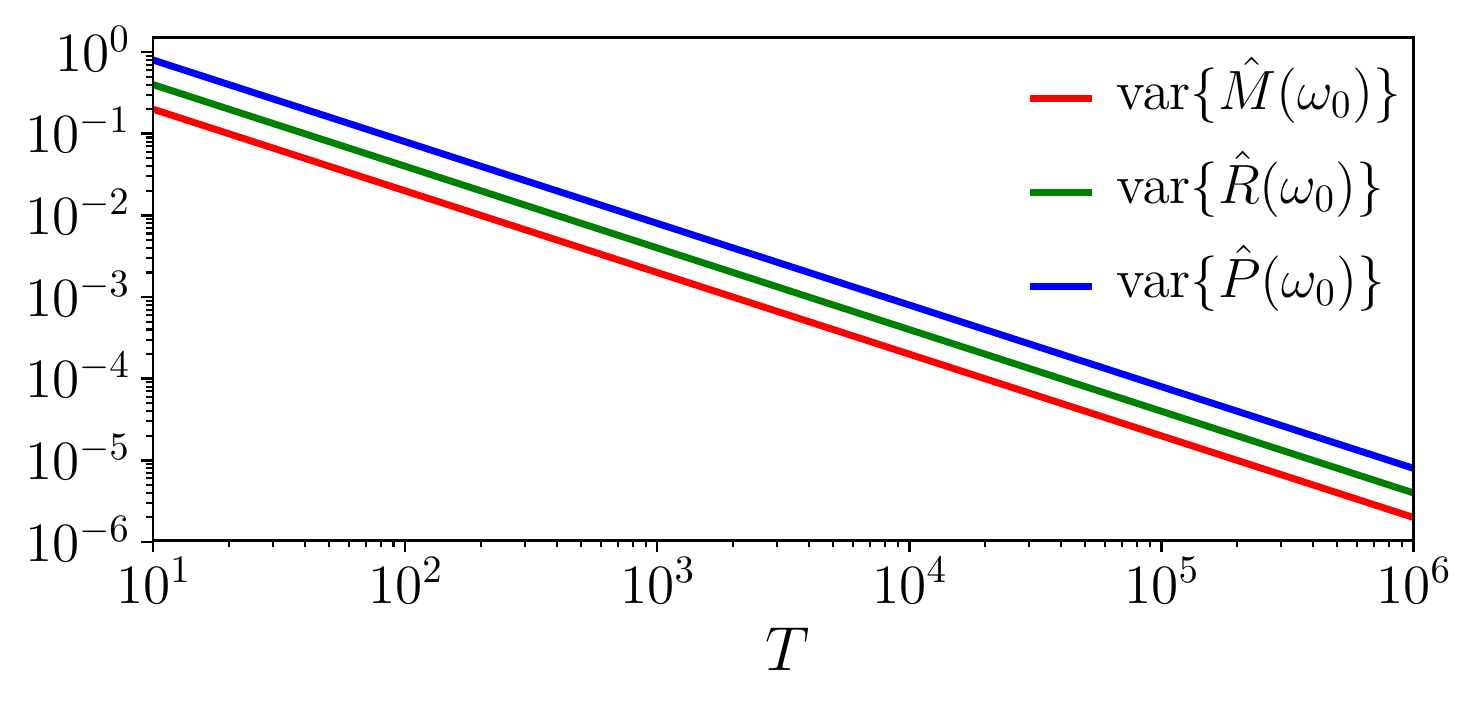} 
		\vspace{-0.3cm}
		\caption{\small The empirical estimation variance as a function of the sample size, $T$, for the proposed spectral moment estimators, computed over 1000 independent realisations.} 
		\label{fig:statistical_consistency}
	\end{figure}

\end{example}



\section{Hypothesis Testing}

\label{sec:GLRT}


The introduced probabilistic framework makes it possible to develop statistical hypothesis tests for the detection of harmonicity, wide-sense stationarity and cyclostationarity -- the subject of this section. This is achieved by introducing a class of generalized likelihood ratio tests (GLRTs), whereby the goodness of fit of two competing statistical models is assessed based on the ratio of their likelihoods.

Consider $T$ samples of the time-spectrum estimate, $\hat{\ubbx}(t) = \uboldPhi^{\Her}(t)\x(t)$, drawn from the distribution in (\ref{eq:pdf_time_spectrum}), denoted by $p(\x,t,\boldtheta)$ with $\boldtheta=\{\ubbm,\ubbR\}$ denoting the set of parameters which characterise the pdf. For a given GLRT, two statistical models are considered: (i) the null hypothesis, $\mathcal{H}_{0}$, which states that the observables are drawn from the density function $p(\x,t,\boldtheta_{0})$; and (ii) the alternative hypothesis, $\mathcal{H}_{1}$, which states that the observables are drawn from the density function $p(\x,t,\boldtheta_{1})$. More formally, we can formulate the test as follows
\begin{align}
	\mathcal{H}_{0} & : \boldtheta_{0}=\{\ubbm_{0},\ubbR_{0}\} \\
	\mathcal{H}_{1} & : \boldtheta_{1}=\{\ubbm_{1},\ubbR_{1}\}
\end{align}
The likelihood of $T$ samples being drawn under either hypothesis, $\mathcal{H}_{i}$ for $i \in \{0,1\}$, based on the least squares estimate of $\ubbx(t) = \uboldPhi^{\Her}(t)\x(t)$, is therefore given by
\begin{align}
	& \mathcal{L}(\x,\boldtheta_{i}) = \prod_{t=0}^{T-1} p(\x,t,\boldtheta_{i}) \notag\\
	& \simeq \frac{ \exp\left[ -\frac{1}{2} \sum_{t=0}^{T-1} \! \left(\uboldPhi^{\Her}(t)\x - \ubbm_{i}\right)^{\!\Her}\!\ubbR_{i}^{-1}\!\left(\uboldPhi^{\Her}(t)\x - \ubbm_{i}\right) \right] }{\pi^{TMN} \, \det^{\frac{T}{2}}\left( \ubbR_{i}\right) } \notag\\
	& = \frac{ \exp\!\left[ -\frac{T}{2} \! \left( \tr{\ubbR_{i}^{-1}\hat{\ubbR}} \! + \! (\hat{\ubbm} \! - \! \ubbm_{i})^{\Her}\ubbR_{i}^{-1}(\hat{\ubbm} \! - \! \ubbm_{i}) \right) \right] }{ \pi^{TMN} \det^{\frac{T}{2}}\!(\ubbR_{i}) } \label{eq:likelihood_GLRT}
\end{align}
where $\hat{\ubbm}$ and $\hat{\ubbR}$ are the empirical approximative ML estimates, defined respectively in (\ref{eq:ML_approx_mean}) and (\ref{eq:ML_approx_cov}). In this case, the GLR decision statistic for the null hypothesis, $\mathcal{H}_{0}$, becomes
\begin{align}
	& \lambda(\x) = - 2 \ln \left( \frac{\underset{\boldtheta_{0}}{\max} \;\; \mathcal{L}(\x,\boldtheta_{0}) }{\underset{\boldtheta_{1}}{\max} \;\; \mathcal{L}(\x,\boldtheta_{1}) } \right) \label{eq:GLRT_ratio}
\end{align}
with the quantity within the $\ln(\cdot)$ operator referred to as the GLR. Note that the maximisers of $\mathcal{L}(\x,\boldtheta_{0})$ and $\mathcal{L}(\x,\boldtheta_{1})$ are given respectively by the ML estimates of the parameters in $\boldtheta_{0}$ and $\boldtheta_{1}$ \cite{Mardia1979}. This procedure is not generally optimal in the Neyman-Pearson sense, but is widely used in practice owing to its reliable performance \cite{Kay1998,Kay2002,Schreier2006,Adali2009}.

Assuming that $\mathcal{H}_{0}$ is true, \textit{Wilk's theorem} \cite{Wilks1938} asserts that the GLR statistic, $\lambda(\x)$, is asymptotically chi-squared distributed with $\nu$ degrees of freedom as the sample size, $T$, increases \cite{Kay1998}, that is
\begin{align}
	\lim_{T \to \infty}	\lambda(\x) \sim \chi_{\nu}^{2}
\end{align}
where $\nu$ is the difference between the cardinalities of the parameter sets, $\boldtheta_{1}$ and $\boldtheta_{0}$. 

The GLRT decision rule is to reject the null hypothesis, $\mathcal{H}_{0}$, if $\lambda(\x)$ exceeds the $\left(1 - \alpha\right)$-th percentile of a chi-squared distribution with $\nu$ degrees of freedom. With this asymptotic result, we can choose a threshold, $\gamma$, which yields a desired probability of false alarm, $P_{\text{FA}}$, to be equal to $\alpha$, that is
\begin{align}
	P_{\text{FA}} = P(\chi_{\nu}^{2}>\gamma) = \alpha
\end{align}
We would therefore reject $\mathcal{H}_{0}$ if $\lambda(\x) > \gamma$. The corresponding probability of rejection, given $\mathcal{H}_{0}$ is true (probability of false alarm), $P_{\text{FA}}$, and the probability of rejection, given $\mathcal{H}_{1}$ is true (probability of detection), $P_{\text{D}}$, are given respectively by
\begin{align}
	P_{\text{FA}} & = \frac{\int_{\lambda(\x)>\gamma} \mathcal{L}(\x,\boldtheta_{0}) \, d\x}{ \int_{\domR^{N}} \mathcal{L}(\x,\boldtheta_{0}) \, d\x } \label{eq:prob_FA} \\
	P_{\text{D}} & = \frac{\int_{\lambda(\x)>\gamma} \mathcal{L}(\x,\boldtheta_{1}) \, d\x}{ \int_{\domR^{N}} \mathcal{L}(\x,\boldtheta_{0}) \, d\x } \label{eq:prob_D}
\end{align}
From the relations in (\ref{eq:likelihood_GLRT}) and (\ref{eq:GLRT_ratio}), the statistic, $\lambda(\x)$, can be expressed in terms of the ML estimates and the parameters of the null and alternative hypotheses, to yield
\begin{align}
	\lambda(\x) & = T \ln \det(\ubbR_{0}) - T \ln \det(\ubbR_{1}) \notag\\
	& \quad + T\,\tr{\ubbR_{0}^{-1}\hat{\ubbR}} \! + \! T \,(\hat{\ubbm} \! - \! \ubbm_{0})^{\Her}\ubbR_{0}^{-1}(\hat{\ubbm} \! - \! \ubbm_{0}) \notag\\
	& \quad - T\,\tr{\ubbR_{1}^{-1}\hat{\ubbR}} \! - \! T \,(\hat{\ubbm} \! - \! \ubbm_{1})^{\Her}\ubbR_{1}^{-1}(\hat{\ubbm} \! - \! \ubbm_{1})
\end{align}
Based on the proposed statistical model of the TFRs, we next introduce a series of hypothesis tests for the detection of harmonicity, cyclostationarity, or general nonstationarity. This is enabled by the results provided in Section \ref{sec:class_nonstationary_processes}, which are summarised in Table \ref{table:nonstationary_classes}.

\begin{table}[h!]
	\centering
	\caption{\small Spectral statistics for special cases of the class of multivariate nonstationary real-valued processes.}
	\label{table:nonstationary_classes}
	\def\arraystretch{1.2}
	\small
	\begin{tabular}{|l|c|c|}
		\hline
		Time-domain behaviour &  $\bbm$ & $\bbR$ and $\bbP$  \\
		\hline
		Harmonicity & $\|\bbm\|>0$ & - \\
		Wide-sense stationary & - & $\bbR = \diag{\bbR}$, $\bbP=\0$  \\ 
		Pure cyclostationary & - & $\|\bbR\|=\|\bbP\|$ \\
		General cyclostationary & - & $\|\bbR\| > \|\bbP\| > 0$ \\
		\hline
	\end{tabular}
\end{table}


\subsection{Test for harmonics}

\label{sec:GLRT_harmonics}

We have shown that harmonics in the time domain are characterised by a non-zero spectral mean, $\ubbm$. We can therefore develop a hypothesis test for the detection of harmonics, whereby the null hypothesis asserts that the spectral mean is zero, while the alternative hypothesis asserts the spectral mean is non-zero, that is
\begin{align}
\mathcal{H}_{0} & : \boldtheta_{0}=\{\0,\ubbR\} \\
\mathcal{H}_{1} & : \boldtheta_{1}=\{\ubbm,\ubbR\}
\end{align}
Because no assumptions are imposed on the spectral covariance, it is assumed to take a general structure in both hypotheses. The GLR statistic in scenario then reduces to
\begin{align}
	\lambda(\x) = T \, \hat{\ubbm}^{\Her}\hat{\ubbR}^{-1}\hat{\ubbm} \label{eq:GLR_stat_harmonic}
\end{align}

\begin{remark}
	Observe that the GLR statistic in (\ref{eq:GLR_stat_harmonic}) implicitly provides us with a measure for quantifying the multivariate signal-to-noise ratio (SNR), that is, $\lambda(\x) = T \, \text{SNR}$. We can therefore employ the following SNR expression
	\begin{align}
		\text{SNR} & = \ubbm^{\Her}\ubbR^{-1}\ubbm \label{eq:multivariate_SNR}
	\end{align}
\end{remark}

%
%

\subsection{Test for cyclostationarity}

\label{sec:GLRT_cyclostationarity}

Cyclostationarity in the time domain is manifested by non-zero dual-frequency spectral covariances or non-zero spectral pseudo-covariances. In other words, cyclostationarity arises when there exists non-zero off-diagonal elements in $\ubbR$. Therefore, the hypothesis test for cyclostationarity detection can be devised analytically, whereby the null hypothesis asserts that the augmented spectral covariance is a diagonal matrix, while the alternative hypothesis asserts the spectrum exhibits non-zero off-diagonal elements, that is
\begin{align}
\mathcal{H}_{0} & : \boldtheta_{0}=\{\ubbm,\diag{\ubbR}\} \\
\mathcal{H}_{1} & : \boldtheta_{1}=\{\ubbm,\ubbR\}
\end{align}
Notice that, because no assumptions are imposed on the spectral mean, it is assumed to take a general structure in both hypotheses. The GLR statistic in this scenario becomes
\begin{align}
	\lambda(\x) & = T\ln\det\left(\diag{\hat{\ubbR}}\right) - T\ln\det\left(\hat{\ubbR}\right) \notag\\
	& \quad + T \, \tr{\diag{\hat{\ubbR}}^{-1}\hat{\ubbR}}  - 2TNM \label{eq:GLR_stat_cyclo}
\end{align}

\begin{remark}
	The GLR statistic in (\ref{eq:GLR_stat_cyclo}) implicitly provides us with new measure for quantifying the \textit{degree of multivariate cyclostationarity}, given by
	\begin{align}
		\varrho & = 1 - \frac{\det(\ubbR)}{\det(\diag{\ubbR})}, \quad 1 \geq \varrho \geq 0
	\end{align}
	which is closely related to the measure proposed in \cite{Schreier2008_3} to quantify the degree of impropriety of multivariate complex variables. Observe that, if $\ubbR$ has at least one non-zero off-diagonal entry (which is the condition for cyclostationarity), then $\varrho > 0$. In the limit case, if $\|\bbP\|=\|\bbR\|$ (the condition for pure cyclostationarity), we then have $\varrho = 1$. Otherwise, if $\ubbR$ is a diagonal matrix (the condition for wide-sense stationarity), then $\varrho = 0$.
\end{remark}

\begin{remark} \label{remark:detection_cyclostationarity}
	Hypothesis tests for the detection of cyclostationarity have already been proposed in \cite{Gardner1990,Giannakis1994,Wisdom2016_2,Wisdom2016,Schreier2014,Schreier2015}, however, these existing techniques are based on the absolute second-order moments of TFRs and therefore cannot distinguish between a harmonic process and a cyclostationary one, since the absolute second-order moments contain both the mean and the centred second-order moments, i.e. $\tilde{\ubbR} = \ubbm\,\ubbm^{\Her}+\ubbR$ (see Remark \ref{remark:PSD}). On the other hand, our approach is based on the centred second-order moments and will therefore only detect cyclostationarity.
\end{remark}

\subsection{Test for general nonstationarity}

\label{sec:GLRT_nonstationarity}

Within the proposed setting, general nonstationarity is defined as the presence of either harmonics or cyclostationarity, or both simultaneously. With that, general nonstationarity in the time domain is manifested by either a non-zero spectral mean and/or by an augmented spectral covariance with non-zero off-diagonal entries. Therefore, the hypothesis test for the detection of general nonstationarity can be established as follows
\begin{align}
\mathcal{H}_{0} & : \boldtheta_{0}=\{\0,\diag{\ubbR}\} \\
\mathcal{H}_{1} & : \boldtheta_{1}=\{\ubbm,\ubbR\}
\end{align}
In this case, the GLR statistic takes the form
\begin{align}
	\lambda(\x) & = T \ln \det\left(\diag{\hat{\ubbR}}\right) - T \ln \det\left(\hat{\ubbR}\right) \notag\\
	& \quad + T\,\tr{\diag{\hat{\ubbR}}^{-1}\left( \hat{\ubbm}\,\hat{\ubbm}^{\Her} + \hat{\ubbR} \right)} \! - 2\,TNM \label{eq:GLR_stat_general_nonstationary}
\end{align}


\begin{remark}
	Following on from Remark \ref{remark:detection_cyclostationarity}, the proposed hypothesis test for the detection of general cyclostationarity is closely related to the existing tests for cyclostationarity which employ the absolute second-order moment, $\tilde{\ubbR} = \ubbm\,\ubbm^{\Her}+\ubbR$. Observe that the absolute second-order moment also appears in the GLR statistic in (\ref{eq:GLR_stat_general_nonstationary}).
\end{remark}



For each of the proposed hypothesis tests, the degrees of freedom parameter, $\nu$, takes the form
\begin{align}
	\nu = \begin{cases}
		NM, & \text{\small test for harmonics},\\
		\frac{1}{2}\left( 3N^{2}M^{2} - NM\right) , & \text{\small test for cyclostationarity},\\
		\frac{1}{2}\left( 3N^{2}M^{2} + NM\right) , & \text{\small test for nonstationarity}.\\
	\end{cases}
\end{align}

\subsection{Receiver operating characteristics (ROC)}

\begin{figure}[h!]
	\centering
	\begin{subfigure}[t]{0.24\textwidth}
		\centering
		\includegraphics[width=1\textwidth, trim={0.2cm 0.2cm 0 0.2cm}, clip]{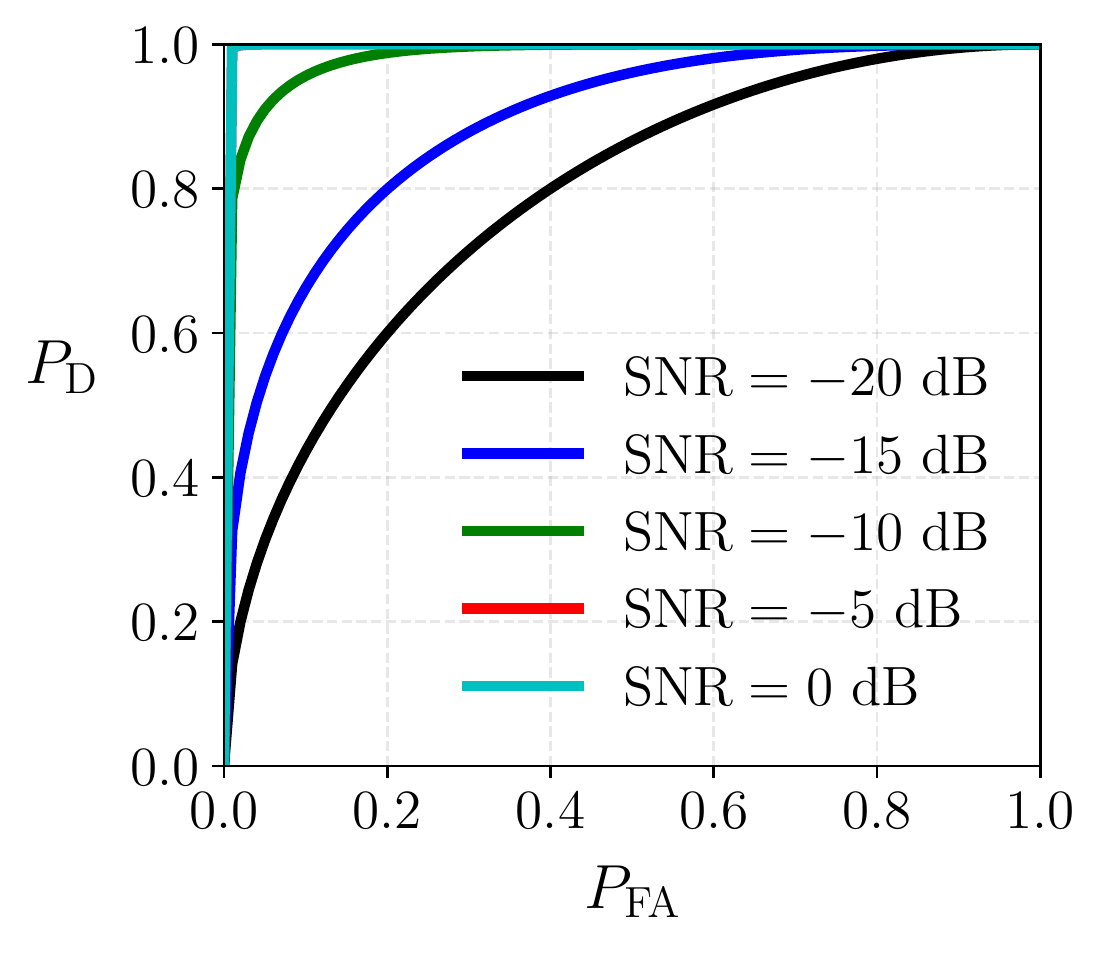} 
		\caption[]%
		{{\small Harmonics detector.}}    
		\label{fig:GLRT_harmonics}
	\end{subfigure}
	\hfill
	\begin{subfigure}[t]{0.24\textwidth}
		\centering
		\includegraphics[width=1\textwidth, trim={0.2cm 0.2cm 0 0.2cm}, clip]{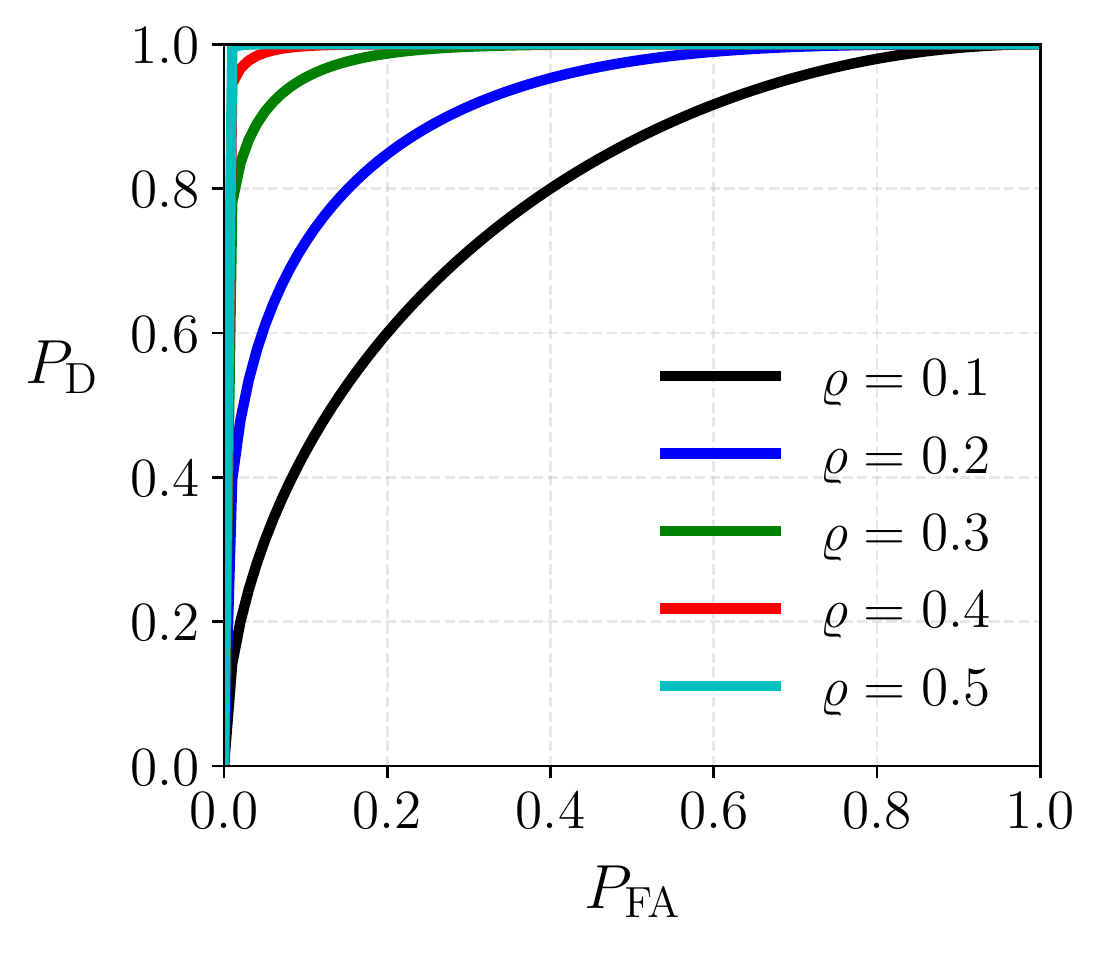} 
		\caption[]%
		{{\small Cyclostationarity detector.}}    
		\label{fig:GLRT_cyclo}
	\end{subfigure}
	\caption[]
	{\small The ROC curves for the proposed hypothesis tests. (a) The ROC curves of the GLRT detector for harmonics proposed in Section \ref{sec:GLRT_harmonics}, evaluated from signals with varying SNRs in the range $[-20,0] \,$dB. (b) The ROC curves of the GLRT detector for cyclostationarity proposed in Section \ref{sec:GLRT_cyclostationarity}, evaluated from signals with varying measures of cyclostationarity, $\varrho$, in the range $[0,0.5]$.} 
	\label{fig:GLRT_single}
\end{figure}

\begin{figure}[h!]
	\centering
	\begin{subfigure}[t]{0.24\textwidth}
		\centering
		\includegraphics[width=1\textwidth, trim={0.2cm 0.2cm 0 0.2cm}, clip]{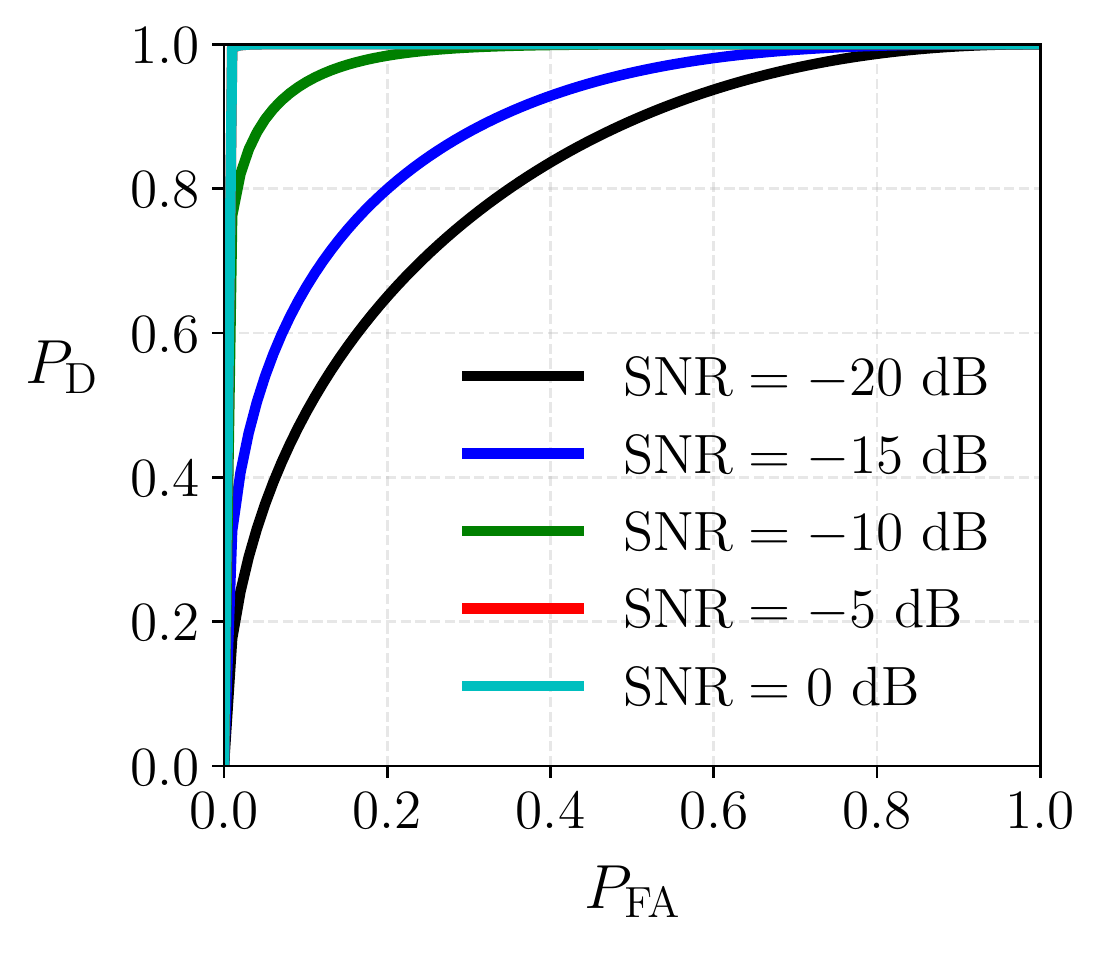} 
		\caption[]%
		{\centering{\small Varying SNR, $\varrho=0$.}}    
		\label{fig:GLRT_nonstationary_harmonics_nocylo}
	\end{subfigure}
	\hfill
	\begin{subfigure}[t]{0.24\textwidth}
		\centering
		\includegraphics[width=1\textwidth, trim={0.2cm 0.2cm 0 0.2cm}, clip]{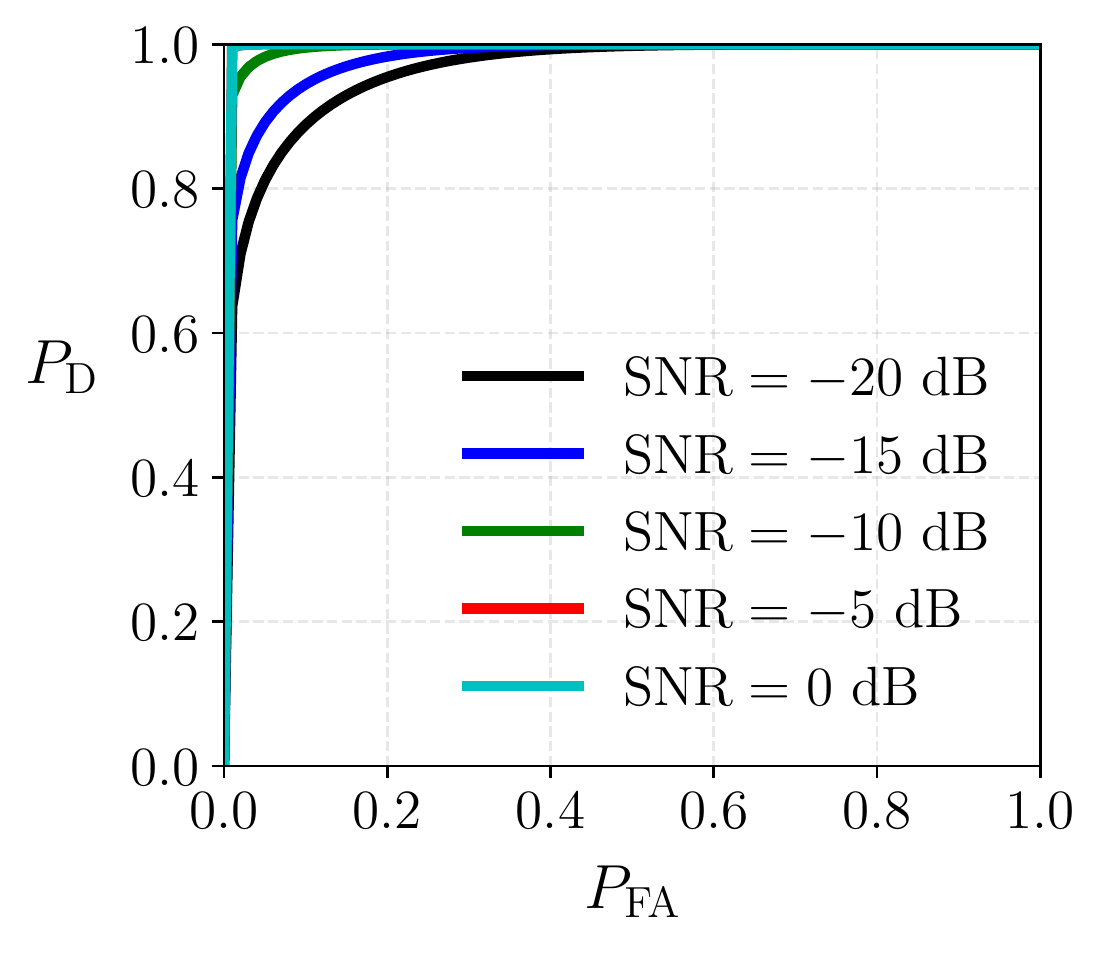} 
		\caption[]%
		{\centering{\small Varying SNR, $\varrho=0.25$.}}    
		\label{fig:GLRT_nonstationary_harmonics_cylo}
	\end{subfigure}
	\vskip\baselineskip
	\centering
	\begin{subfigure}[t]{0.24\textwidth}
		\centering
		\includegraphics[width=1\textwidth, trim={0.2cm 0.2cm 0 0.2cm}, clip]{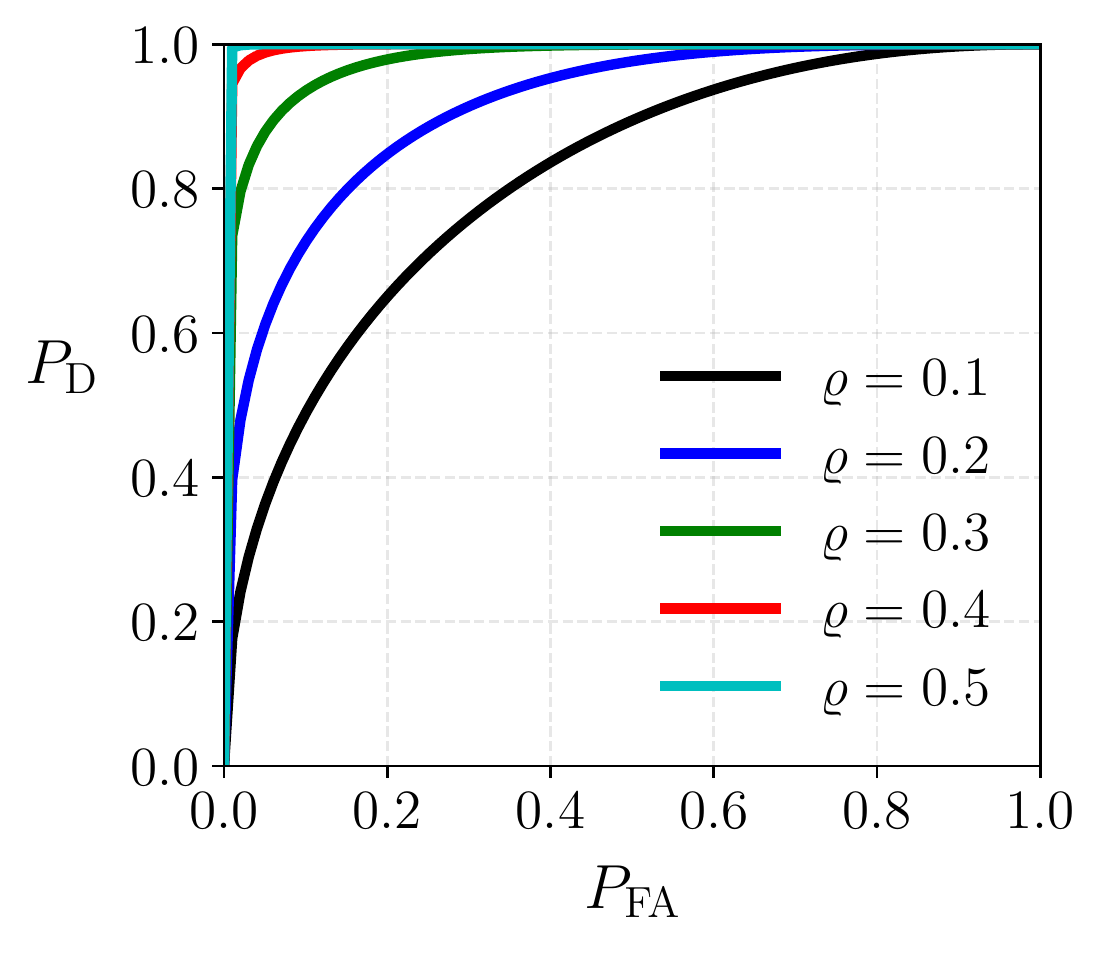} 
		\caption[]%
		{\centering{\small Varying $\varrho$, $\text{SNR}=0$.}}    
		\label{fig:GLRT_nonstationarity_cyclo_noharmonics}
	\end{subfigure}
	\hfill
	\begin{subfigure}[t]{0.24\textwidth}  
		\centering 
		\includegraphics[width=1\textwidth, trim={0.2cm 0.2cm 0 0.2cm}, clip]{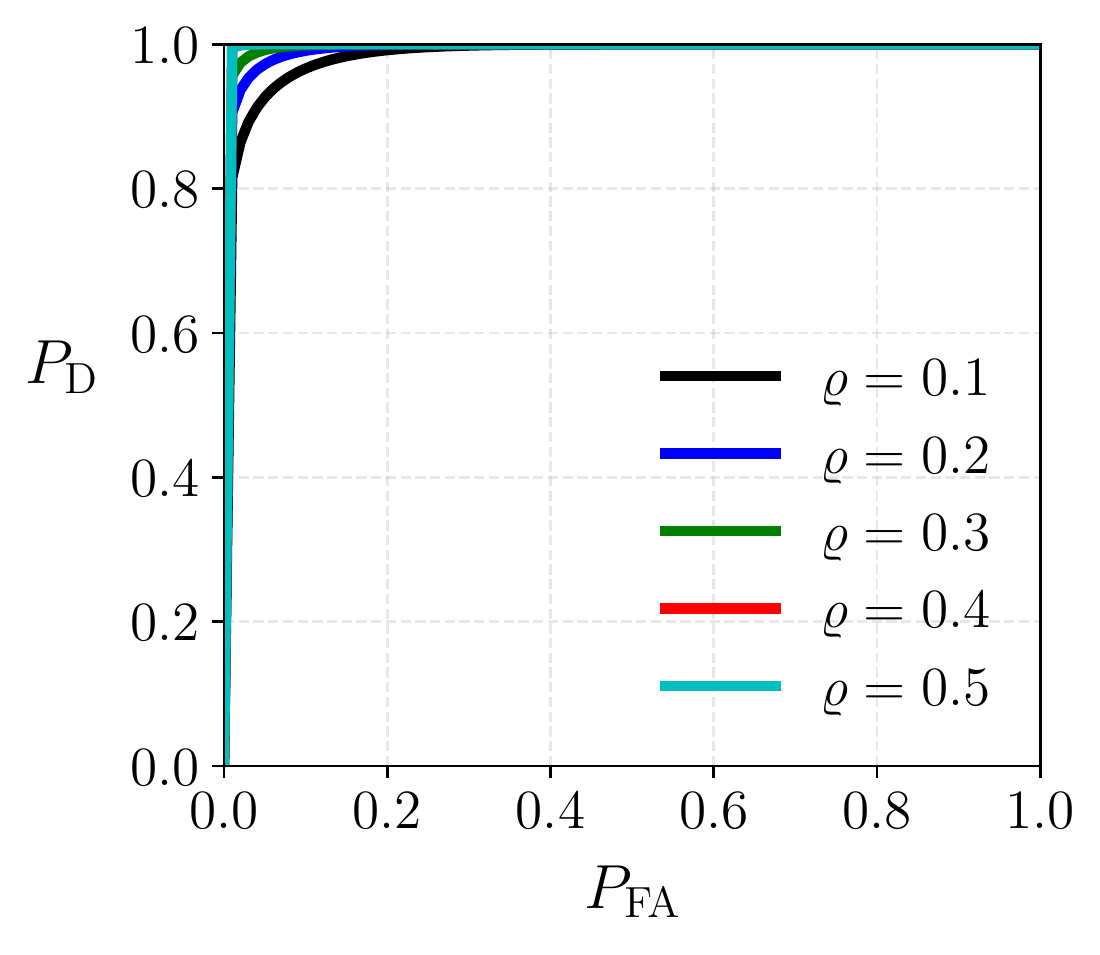} 
		\caption[]%
		{\centering{\small Varying $\varrho$, $\text{SNR}=-10 \,$dB.}}    
		\label{fig:GLRT_nonstationarity_cyclo_harmonics}
	\end{subfigure}
	\caption[]
	{\small The ROC curves for the proposed hypothesis tests. (a) The ROC curves of the GLRT detector for general nonstationarity proposed in Section \ref{sec:GLRT_nonstationarity}. (a) ROC curves evaluated from signals with varying SNRs in the range $[-20,0] \,$dB in the absence of cyclostationarity, $\varrho=0$. (b) ROC curves evaluated from signals with varying SNRs in the range $[-20,0] \,$dB in the presence of cyclostationarity, $\varrho=0.25$. (c) ROC curves evaluated from signals with varying $\varrho$ in the range $[0,0.5]$ in the absence of harmonics, $\bbm=\0$. (d) ROC curves evaluated from signals with varying $\varrho$ in the range $[0,0.5]$ in the presence of harmonics, $\text{SNR}=-10 \,$dB.} 
	\label{fig:GLRT_combo}
\end{figure}

From the definitions of $P_{\text{FA}}$ and $P_{\text{D}}$ in (\ref{eq:prob_FA})-(\ref{eq:prob_D}), an increase in the decision threshold, $\gamma$, will result in a decrease in both the probabilities of false alarm and detection. We therefore investigated the achievable $(P_{\text{D}}, P_{\text{FA}})$ pairs for each of the proposed hypothesis tests by varying the value of $\gamma$ on the so called \textit{receiver operating characteristic} (ROC) curves. The ROC curves of the GLRT detectors for harmonics, cyclostationarity and general nonstationarity are respectively shown in Fig. \ref{fig:GLRT_harmonics}, Fig. \ref{fig:GLRT_cyclo} and Fig. \ref{fig:GLRT_combo}.

For each of the hypothesis tests, the probabilities, $P_{\text{FA}}$ and $P_{D}$, of the associated GLRT detector were evaluated over $1000$ independent Monte Carlo simulations using the sampling procedure given in Section \ref{sec:sampling_procedure}. The dimensionality of the multivariate nonstationary process, $\x(t) \in \domR^{N}$, was set to $N=10$ and the number of samples in each signal realisation was $T = 500$.  The spectral moments of the sampled signal were chosen arbitrarily to be non-zero, that is, $\|\bbm\|>0$, $\|\bbR\|>0$ and $\|\bbP\|>0$, unless stated otherwise.


\section{Conclusions}

A class of multivariate spectral representations of real-valued nonstationary processes has been introduced. This has been achieved based on their time-frequency representation (TFR) being general complex Gaussian distributed. In this way, various nonstationary time-domain signal properties have been naturally parametrized by the first- and second-order moments of the TFRs. The maximum likelihood estimator of these spectral moments has also been derived and employed within a GLRT for nonstationarity detection. Simulation examples have demonstrated the benefits of the proposed framework over existing ones for spectral analysis in low-SNR environments.

\pagebreak

\footnotesize

\bibliographystyle{IEEEtran}
\bibliography{./Bibliography} 

\end{document}